\documentclass[amsmath,notitlepage,amssymb,aps,showkeys,floatfix,prd,a4paper,
  twocolumn,nofootinbib]{revtex4-1}

\usepackage{graphicx}
\usepackage{amsmath}
\usepackage{epstopdf}
\usepackage{amsfonts,amssymb}
\usepackage[utf8]{inputenc}
\usepackage{pstricks}
\usepackage{color}

\usepackage[compat=1.0.0]{tikz-feynman}
\usepackage{simplewick}
\usepackage{color, colortbl}
\definecolor{Gray}{gray}{0.9}
\usepackage{float}
\usepackage{tikz-feynman}
\usepackage{feynmp}
\usepackage{forest}

\usepackage{verbatim}    
\usepackage{dcolumn}
\usepackage{bm}
\usepackage{epsfig}
\usepackage{braket}
\usepackage[normalem]{ulem} 
\usepackage{longtable} 

\def\beq{\begin{equation}}
\def\eeq{\end{equation}}
\def\bea{\begin{eqnarray}}
\def\eea{\end{eqnarray}}
\def\eq{\end{quote}}

\def\nnb{\nonumber}

\def\nnb{\nonumber}
\def\la{\langle}
\def\ra{\rangle}

\def\ba{\vspace*{-0.2cm}\begin{array}}
\def\ea{\end{array}\vspace*{-0.2cm}}

\def\als{\alpha_s}

\def\gg2{ \la\alpha_s G2 \ra}
\def\gg3{g^3f_{abc}\la G^aG^bG^c \ra}
\def\ggg4{\la\als^2G4\ra}

\def\enq{\end{equation}}
\def\beqa{\begin{eqnarray}}
\def\enqa{\end{eqnarray}}
\def\nnb{\nonumber}

\def\MeV{\nobreak\,\mbox{MeV}}
\def\GeV{\nobreak\,\mbox{GeV}}

\def\qq{\lag\bar{q}q\rag}
\def\pli{p^\prime}

\def\si{\sigma}

\newcommand{\mix}{\lag\bar{q}g\si.Gq\rag}
\newcommand{\be}{\begin{equation}}
\newcommand{\ee}{\end{equation}}
\newcommand{\ben}{\begin{eqnarray}}
\newcommand{\een}{\end{eqnarray}}
\newcommand{\lb}{\label}
\def\MeV{\mbox{ MeV}} 
\def\GeV{\mbox{ GeV}}

\newcommand{\rag}{\rangle}
\newcommand{\lag}{\langle}

\def\MeV{\mbox{ MeV}} 
\def\GeV{\mbox{ GeV}} 
\def\mb{\mbox{ mb}} 
\newcommand{\pslash}{\not{\hbox{\kern-2.3pt $p$}}}
\newcommand{\pdslash}{\not{\hbox{\kern-2pt $\partial$}}}

\begin{document}


\title{Interactions of the doubly charmed state $T_{cc} ^+$ with a
  hadronic medium}

\author{ L. M. Abreu}
\email{luciano.abreu@ufba.br}
\affiliation{ Instituto de F\'isica, Universidade Federal da Bahia,
Campus Universit\'ario de Ondina, 40170-115, Bahia, Brazil}
\author{F. S. Navarra}
\email{navarra@if.usp.br}
\affiliation{Instituto de F\'{\i}sica, Universidade de S\~{a}o Paulo, 
Rua do Mat\~ao, 1371, CEP 05508-090,  S\~{a}o Paulo, SP, Brazil}
\author{H. P. L. Vieira}
\email{hilde\underline{ }son@hotmail.com}
\affiliation{ Instituto de F\'isica, Universidade Federal da Bahia,
Campus Universit\'ario de Ondina, 40170-115, Bahia, Brazil}
\author{M. Nielsen}
\email{mnielsen@if.usp.br}
\affiliation{Instituto de F\'{\i}sica, Universidade de S\~{a}o Paulo,
Rua do Mat\~ao, 1371, CEP 05508-090,  S\~{a}o Paulo, SP, Brazil}

\begin{abstract}


We investigate the absorption and
production processes of this new state in a hadronic medium, considering 
the reactions $T_{cc}^+ \pi,  T_{cc}^+ \rho \rightarrow D^{(*)} D^{(*)} $ and
the corresponding inverse reactions.  
We use effective field Lagrangians to account for the couplings between light
and heavy mesons, and give  special attention to the form factors in the
vertices. We calculate here for the first
time the $ T_{cc}^+ - D - D^*$ form factor derived from QCD sum rules. The results are also obtained by testing widely utilized empirical form factors.
The absorption cross sections are
found to be larger than the production ones. We compare our results with the 
only existing  estimate of these quantities, presented in a work of
J.~Hong, S.~Cho, T.~Song and S.~H.~Lee, in which the authors employed the
quasi-free approximation. We find cross sections which are one order of
magnitude smaller. 

\end{abstract}

\maketitle


\section{Introduction}

\label{Introduction}


Very recenlty the LHCb collaboration has reported the observation of a 
narrow peak in the $D^0 D^0 \pi^+ $-mass spectrum in proton-proton $(pp)$
collisions with statistical significance of more than
$10 \, \sigma$~\cite{LHCb:2021vvq,LHCb:2021auc}. By using an amplitude 
model based on the Breit-Wigner formalism, this peak has been fitted to    
one resonance with a mass of approximately $3875 \MeV$ and quantum numbers
$J^P = 1^+$.
Its minimum valence quark content is $c c \bar{u}\bar{d} $, giving it the 
unequivocal status of the first observed unconventional hadron with two    
heavy quarks of the same flavor. According to the data, its binding energy 
with respect to the $D^{*+} D^0 $ mass threshold is        
$273\pm 61 \pm 5 _{-14}^{+11}$ keV and the decay width is              
$410\pm 165 \pm 43 _{-38}^{+18}$ keV. These values are consistent with      
the expected properties for a $T_{cc}^+ $ isoscalar tetraquark ground state
with $J^P = 1^+$. 

Even before the experimental discovery of this doubly charmed tetraquark
state, there was a debate concerning its fundamental aspects, such as its
decay/formation mechanisms and underlying structure
~\cite{Gelman:2002wf,Janc:2004qn,Vijande:2003ki,Navarra:2007yw,
Vijande:2007rf,Ebert:2007rn,Lee:2009rt,Yang:2009zzp,Hong:2018mpk,    
Hudspith:2020tdf,Cheng:2020wxa,Qin:2020zlg}. Its discovery obviously 
stimulated the appearance of more studies  employing different theoretical
approaches~\cite{Agaev:2021vur,Dong:2021bvy, 
Agaev:2021vur,Dong:2021bvy,Huang:2021urd,Li:2021zbw,Ren:2021dsi,Xin:2021wcr, 
Yang:2021zhe,Meng:2021jnw,Ling:2021bir,Fleming:2021wmk,Jin:2021cxj, 
Azizi:2021aib,Hu:2021gdg}. In particular, several works have investigated  the
implications of the $T_{cc}^+ $ structure (hadron molecule or compact
tetraquark) for the observables. However, a compelling understanding of the
nature of the  $T_{cc}^+ $ is still lacking.

In order to determine the internal structure of the $T_{cc}^+ $ state, more
detailed experimental data and theoretical studies are necessary. 
In this context, heavy-ion collisions appear as a promising environment,
where charm quarks are copiously produced. The search for exotic charm hadrons
in heavy-ion collisions has already started and the $X(3872)$ has been observed
by the CMS and LHCb collaborations. In 
these collisions there is a phase transition from nuclear matter to the    
quark-gluon plasma (QGP), i.e. the locally thermalized state of deconfined   
quarks and gluons. The QGP expands, cools down and hadronizes, forming a gas
of hadrons. When this last transition takes place, heavy quarks coalesce to    
form multiquark bound states at the end of the QGP phase. Next, the multiquark
states interact with other hadrons in the course of the hadronic phase. They
can be destroyed in collisions with the comoving light mesons, or produced
through the inverse processes~
\cite{Chen:2007zp,ChoLee1,XProd1,XProd2,UFBaUSP1,MartinezTorres:2017eio,   
  Abreu:2017cof,Abreu:2018mnc,LeRoux:2021adw}.  The final yields depend on
the hadronic interactions which, in turn, depend on the spatial configuration 
of the multiquark systems. Therefore, the evaluation of the interaction cross 
sections of the $T_{cc}$ with light mesons is a crucial ingredient for the
interpretation of the data. While the hadronic interactions of the $X(3872)$
have been addressed in several papers, there is only one work,
\cite{Hong:2018mpk}, where the $T_{cc}$ - light meson cross section was
calculated. In Ref.~\cite{Hong:2018mpk} the authors treated the $T_{cc}$ as a
loosely bound state of a $D$ and a $D^*$. In this approach it seems natural to
use the quasi-free approximation, in which the charm mesons are taken to be
on-shell and their mutual interaction and binding energy are neglected. In the
quasi-free approximation the $T_{cc}$ is absorbed when a pion from hadron gas
interacts with the $D$ or with the $D^*$. In each of these interactions the
other heavy meson is a spectator. The advantage of this approach is that    
the only dynamical ingredient is the $D^* \, D \, \pi$ Lagrangian, which is
well-known.
On the other hand, the role of the quantum numbers of the $D^* \, D$ bound
state is neglected. Moreover, some possible final states are not included.
Clearly the subject deserves further investigation and this is the main
purpose of this work.

It is important to emphasize that in the long time limit,
the collisions mentioned above will drive the exotic charm mesons to chemical
equilibrium, at which point, the only relevant parameters are the particle
mass, the temperature and the charm fugacity. In equilibrium, we can neglect
the microscopic dynamics and calculate the particle abundances with the
statistical hadronization model (SHM) \cite{shm1,shm2}. This model reproduces
very well  most of the hadron multiplicities. For exotic particles  there are no
multiplicity measurements yet. If they are produced by quark coalescence, their
yields in the beginning of the hadron gas phase can be very different from
the equilibrium values. Whether or not equilibrium will be reached in the
fireball lifetime, depends on the microscopic cross sections. In
\cite{LeRoux:2021adw}, for example, it was shown that, in the case of the $K^*$, 
the equilibration time could change by a factor 2 for different choices of the
cross sections. This motivates us to  study the microscopic
dynamics of $T^+_{cc}$ production.


In what follows we will  evaluate the hadronic effects on the  
$T_{cc}^+ $ state. The processes involving its suppression by the interaction with light mesons, as well as their inverse (production) ones, are investigated within an effective approach, looking especially the issue of the form factors in the couplings.


The paper is organized as follows. In Section~\ref{Framework} we describe  
the effective formalism.  
Section~\ref{FormFactors} is devoted to the theory of
form factors and their functional form employed in this work.  
In  Section~\ref{Results} 
we present and analyze the results obtained.  Finally,                 
Section~\ref{Conclusions} is devoted to the concluding remarks.


\section{Framework}
\label{Framework}

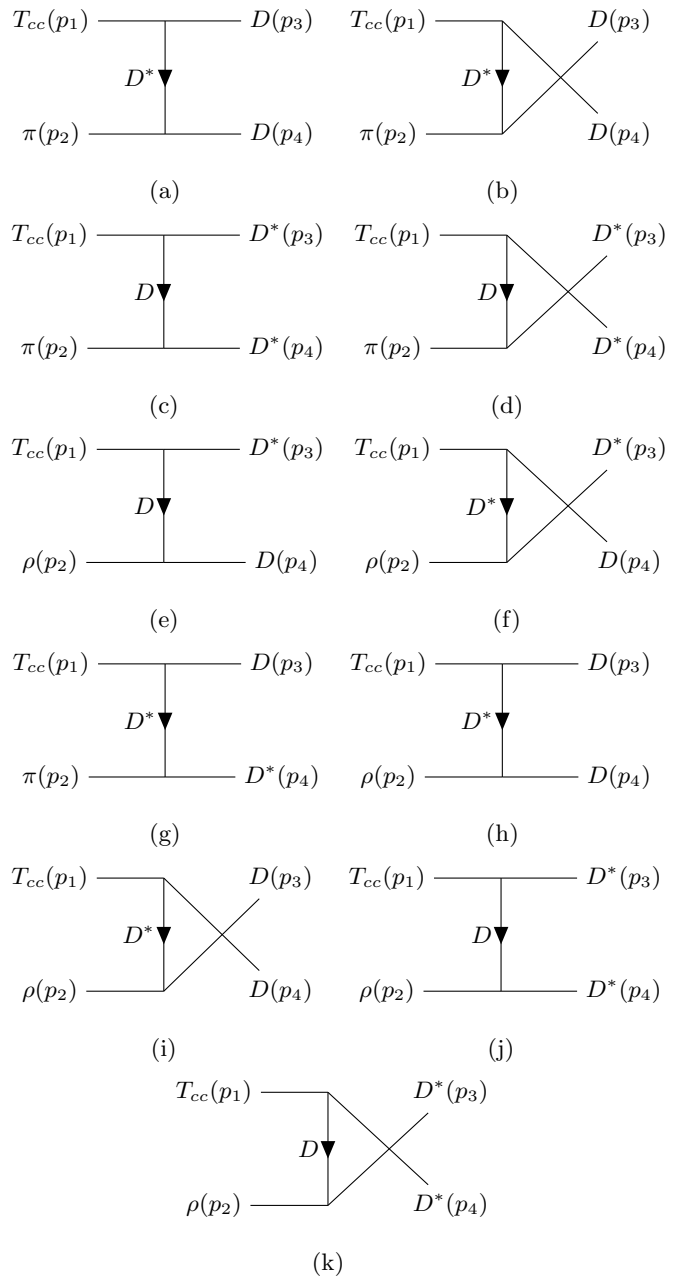
\begin{figure}[!ht]
    \centering

\begin{tikzpicture}
\begin{feynman}
\vertex (a1) {$T_{cc} (p_{1})$};
	\vertex[right=1.5cm of a1] (a2);
	\vertex[right=1.cm of a2] (a3) {$D (p_{3})$};
	\vertex[right=1.4cm of a3] (a4) {$T_{cc} (p_{1})$};
	\vertex[right=1.5cm of a4] (a5);
	\vertex[right=1.cm of a5] (a6) {$D (p_{3})$};
\vertex[below=1.5cm of a1] (c1) {$\pi (p_{2})$};
\vertex[below=1.5cm of a2] (c2);
\vertex[below=1.5cm of a3] (c3) {$D (p_{4})$};
\vertex[below=1.5cm of a4] (c4) {$\pi (p_{2})$};
\vertex[below=1.5cm of a5] (c5);
\vertex[below=1.5cm of a6] (c6) {$D (p_{4})$};
	\vertex[below=2cm of a2] (d2) {(a)};
	\vertex[below=2cm of a5] (d5) {(b)};
\diagram* {
(a1) -- (a2), (a2) -- (a3), (c1) -- (c2), (c2) -- (c3), (a2) -- [fermion, edge label'= $D^{*}$] (c2), (a4) -- (a5), (a5) -- (c6), (c4) -- (c5), (c5) -- (a6), (a5) -- [fermion, edge label'= $D^{*}$] (c5)
}; 
\end{feynman}
\end{tikzpicture}

\begin{tikzpicture}
\begin{feynman}
\vertex (a1) {$T_{cc} (p_{1})$};
	\vertex[right=1.5cm of a1] (a2);
	\vertex[right=1.cm of a2] (a3) {$D^* (p_{3})$};
	\vertex[right=1.4cm of a3] (a4) {$T_{cc}  (p_{1})$};
	\vertex[right=1.5cm of a4] (a5);
	\vertex[right=1.cm of a5] (a6) {$D^* (p_{3})$};
\vertex[below=1.5cm of a1] (c1) {$\pi (p_{2})$};
\vertex[below=1.5cm of a2] (c2);
\vertex[below=1.5cm of a3] (c3) {$D^* (p_{4})$};
\vertex[below=1.5cm of a4] (c4) {$\pi (p_{2})$};
\vertex[below=1.5cm of a5] (c5);
\vertex[below=1.5cm of a6] (c6) {$D^* (p_{4})$};
	\vertex[below=2cm of a2] (d2) {(c)};
	\vertex[below=2cm of a5] (d5) {(d)};
\diagram* {
(a1) -- (a2), (a2) -- (a3), (c1) -- (c2), (c2) -- (c3), (a2) -- [fermion, edge label'= $D$] (c2), (a4) -- (a5), (a5) -- (c6), (c4) -- (c5), (c5) -- (a6), (a5) -- [fermion, edge label'= $D$] (c5)
}; 
\end{feynman}
\end{tikzpicture}

\begin{tikzpicture}
\begin{feynman}
\vertex (a1) {$T_{cc} (p_{1})$};
	\vertex[right=1.5cm of a1] (a2);
	\vertex[right=1.cm of a2] (a3) {$D^* (p_{3})$};
	\vertex[right=1.4cm of a3] (a4) {$T_{cc} (p_{1})$};
	\vertex[right=1.5cm of a4] (a5);
	\vertex[right=1.cm of a5] (a6) {$D^* (p_{3})$};
\vertex[below=1.5cm of a1] (c1) {$\rho (p_{2})$};
\vertex[below=1.5cm of a2] (c2);
\vertex[below=1.5cm of a3] (c3) {$D (p_{4})$};
\vertex[below=1.5cm of a4] (c4) {$\rho (p_{2})$};
\vertex[below=1.5cm of a5] (c5);
\vertex[below=1.5cm of a6] (c6) {$D (p_{4})$};
	\vertex[below=2cm of a2] (d2) {(e)};
	\vertex[below=2cm of a5] (d5) {(f)};
\diagram* {
(a1) -- (a2), (a2) -- (a3), (c1) -- (c2), (c2) -- (c3), (a2) -- [fermion, edge label'= $D$] (c2), (a4) -- (a5), (a5) -- (c6), (c4) -- (c5), (c5) -- (a6), (a5) -- [fermion, edge label'= $D^{*}$] (c5)
}; 
\end{feynman}
\end{tikzpicture}

\begin{tikzpicture}
\begin{feynman}
\vertex (a1) {$T_{cc} (p_{1})$};
	\vertex[right=1.5cm of a1] (a2);
	\vertex[right=1.cm of a2] (a3) {$D (p_{3})$};
	\vertex[right=1.4cm of a3] (a4) {$T_{cc} (p_{1})$};
	\vertex[right=1.5cm of a4] (a5);
	\vertex[right=1.cm of a5] (a6) {$D (p_{3})$};
\vertex[below=1.5cm of a1] (c1) {$\pi (p_{2})$};
\vertex[below=1.5cm of a2] (c2);
\vertex[below=1.5cm of a3] (c3) {$D^* (p_{4})$};
\vertex[below=1.5cm of a4] (c4) {$\rho (p_{2})$};
\vertex[below=1.5cm of a5] (c5);
\vertex[below=1.5cm of a6] (c6) {$D (p_{4})$};
	\vertex[below=2cm of a2] (d2) {(g)};
	\vertex[below=2cm of a5] (d5) {(h)};
\diagram* {
(a1) -- (a2), (a2) -- (a3), (c1) -- (c2), (c2) -- (c3), (a2) -- [fermion, edge label'= $D^*$] (c2),(a4) -- (a5), (a5) -- (a6), (c4) -- (c5), (c5) -- (c6), (a5) -- [fermion, edge label'= $D^*$] (c5)
}; 
\end{feynman}
\end{tikzpicture}

\begin{tikzpicture}
\begin{feynman}
\vertex (a1) {$T_{cc} (p_{1})$};
	\vertex[right=1.5cm of a1] (a2);
	\vertex[right=1.cm of a2] (a3) {$D (p_{3})$};
	\vertex[right=1.4cm of a3] (a4) {$T_{cc} (p_{1})$};
	\vertex[right=1.5cm of a4] (a5);
	\vertex[right=1.cm of a5] (a6) {$D^{*} (p_{3})$};
\vertex[below=1.5cm of a1] (c1) {$\rho (p_{2})$};
\vertex[below=1.5cm of a2] (c2);
\vertex[below=1.5cm of a3] (c3) {$D (p_{4})$};
\vertex[below=1.5cm of a4] (c4) {$\rho (p_{2})$};
\vertex[below=1.5cm of a5] (c5);
\vertex[below=1.5cm of a6] (c6) {$D^{*} (p_{4})$};
	\vertex[below=2cm of a2] (d2) {(i)};
	\vertex[below=2cm of a5] (d5) {(j)};

\diagram* {
(a1) -- (a2), (a2) -- (c3), (c1) -- (c2), (c2) -- (a3), (a2) -- [fermion, edge label'= $D^*$] (c2), (a4) -- (a5), (a5) -- (a6), (c4) -- (c5), (c5) -- (c6), (a5) -- [fermion, edge label'= $D$] (c5)
}; 
\end{feynman}
\end{tikzpicture}

\begin{tikzpicture}
\begin{feynman}
\vertex (a1) {$T_{cc} (p_{1})$};
	\vertex[right=1.5cm of a1] (a2);
	\vertex[right=1.cm of a2] (a3) {$D^* (p_{3})$};
\vertex[below=1.5cm of a1] (c1) {$\rho (p_{2})$};
\vertex[below=1.5cm of a2] (c2);
\vertex[below=1.5cm of a3] (c3) {$D^* (p_{4})$};
	\vertex[below=2cm of a2] (d2) {(k)};
\diagram* {
(a1) -- (a2), (a2) -- (c3), (c1) -- (c2), (c2) -- (a3), (a2) -- [fermion, edge label'= $D$] (c2)
}; 
\end{feynman}
\end{tikzpicture}
 
        \caption{Diagrams contributing to the following process (without specification of the charges of the particles): $ T_{cc} \pi \rightarrow D D $  [(a) and (b)], $ T_{cc} \pi \rightarrow D^* D^* $  [(c) and (d)], $ T_{cc} \rho \rightarrow D^* D $  [(e) and (f)], $ T_{cc} \pi \rightarrow D D^* $  [(g)], $ T_{cc} \rho \rightarrow D D $  [(h) and (i)] and $ T_{cc} \rho \rightarrow D^* D^* $  [(j) and (k)]. The particle charges are not specified. }
\label{DIAG1}
\end{figure}

We are interested in the interactions of the $T_{cc}^+$ state with the
surrounding hadronic medium composed of the lightest pseudoscalar and
vector mesons, i.e. $\pi$ and $\rho$ mesons, respectively. Because of their
large multiplicity (mainly the pions) with respect to other light hadrons,  
the reactions involving them are expected to provide the main contributions. 
In Fig.~\ref{DIAG1} we show the lowest-order Born diagrams
contributing to each process, without the specification of the particle charges.



To calculate the respective cross sections related to the reactions in
Fig.~\ref{DIAG1}, we make use of the effective hadron Lagrangian approach.
Accordingly, for the diagrams $(a)-(f)$ we employ the effective Lagrangians
involving $\pi$, $\rho$,  $D$  and $D^*$ mesons given
by~\cite{Chen:2007zp,ChoLee1,XProd1,XProd2,UFBaUSP1}, 
\begin{eqnarray}\label{Lagr1}
  {\mathcal{L}}_{\pi D D^* } &=& i g_{\pi D D^*} D_\mu ^* \vec{\tau} \cdot
  \left(  \bar{D} \partial^\mu
\vec{\pi} -   \partial^\mu \bar{D} \vec{\pi} \right) + h. c., 
\nonumber\\
\mathcal{L}_{\rho DD} &=&  ig_{\rho DD} (D \vec{\tau } \partial_{\mu} \bar{D}
- \partial_{\mu} D \vec{\tau} \bar{D}) \cdot \vec{\rho}^{\mu}, 
\nonumber\\
\mathcal{L}_{\rho D^* D^*} &=& i g_{\rho D^* D^*} \left[
  (\partial_{\mu} D^{* \nu} \vec{\tau } \bar{D}^{*}_{\nu} - D^{* \nu}
  \vec{\tau } \partial_{\mu} \bar{D}^{*}_{\nu}) \cdot \vec{\rho}^{\mu} \right. 
\nonumber \\
\nonumber  & &+ (D^{* \nu} \vec{\tau } \cdot \partial_{\mu} \vec{\rho}_{\nu}
- \partial_{\mu} D^{* \nu} \vec{\tau } \cdot \vec{\rho}_{\nu}) \bar{D}^{* \mu} \\
& & + \left. D^{*\mu} (\vec{\tau } \cdot \vec{\rho}^{\nu}
\partial_{\mu} \bar{D}^{*}_{\nu} - \vec{\tau } \cdot \partial_{\mu}
\vec{\rho}^{\nu} \bar{D}^{*}_{\nu})\right] ,
\end{eqnarray}
where $\vec{\tau}$ are the Pauli matrices in the isospin space; $\vec{\pi}$
and $\vec{\rho}$ denote the pion and $\rho$-meson isospin triplets; and
$D^{(\ast)} = (D^{(\ast) 0}, D^{(\ast) +}  ) $ represents the isospin
doublet for the pseudoscalar (vector) $D^{(\ast) }$  meson. The coupling
constants $g_{\pi D D^*}, g_{\rho DD}  $ and $  g_{\rho D^{*} D^{*}} $ are
determined from the decay width of $D^{\ast }$ and from the relevant symmetries,
having the following values~\cite{Chen:2007zp,ChoLee1}:  $g_{\pi D D^*} = 6.3$
and $g_{\rho DD} = g_{\rho D^{*} D^{*}} = 2.52$.   

In the case of the diagrams $(g)-(k)$ in Fig.~\ref{DIAG1}, the vertices    
involving light and heavy-light mesons are anomalous, and can be described 
in terms of a gauged Wess-Zumino action~\cite{psipi-oh}. Explicitly,  they
are 
\begin{eqnarray}\label{Lagr2}
	\mathcal{L}_{\pi D^{*} D^{*}} &=&  -g_{\pi D^{*} D^{*}} \varepsilon^{\mu \nu \alpha \beta} \partial_{\mu} D_{\nu}^{*} \pi \partial_{\alpha} \bar{D}_{\beta}^{*}, 
\nonumber \\
\mathcal{L}_{\rho D D^{*}} &=&  -g_{\rho D D^{*}}
\varepsilon^{\mu \nu \alpha \beta} ( D \partial_{\mu} \rho_{\nu} \partial_{\alpha} \bar{D}_{\beta}^{*} + \partial_{\mu} D_{\nu}^{*} \partial_{\alpha} \rho_{\beta} \bar{D} ),\nonumber \\
\end{eqnarray}
with $\epsilon_{0123} = +1$. The coupling constants $g_{\pi D^{*} D^{*}}$ and 
$g_{\rho D D^{*}}$ have the following values~\cite{psipi-oh}: 
$g_{\pi D^{*} D^{*}} = 9.08  \GeV ^{-1}$ and
$g_{\rho D D^{*}} = 2.82  \GeV ^{-1}$.

In this study, we assume that the  $T_{cc}^+$ is a bound state of $ D^* D $, 
with quantum numbers $I (J^P) = 0 (1^+) $. 
Therefore, the effective Lagrangian describing the interaction between the 
$T_{cc}$ and the $D D^*$ pair is given by~\cite{Ling:2021bir},   
\begin{eqnarray}\label{Lagr3}
\mathcal{L}_{T_{cc}} &=& ig_{T_{cc} D D^*} T_{cc}^{\mu}  D_{\mu}^{*} D.
\end{eqnarray}
In the expression above, $T_{cc}$ denotes the field associated to  
$T_{cc}^+$ state; this notation will be used henceforth. Also, the       
$D_{\mu}^{*}  D $ means the $D_{\mu}^{*+} D^0 $ and  $D_{\mu}^{*0} D^+$ 
components, although we do not distinguish them here since we will use   
isospin-averaged masses.  The coupling constant 
$g_{T_{cc} D D^*} $ will be discussed in the next section.


The effective Lagrangians introduced above allow us to determine the
amplitudes of the processes shown in Fig.~\ref{DIAG1}. They are given by
\begin{eqnarray}
\mathcal{M}_{T_{cc} \pi \rightarrow DD} &= &  \mathcal{M}_{T_{cc}^{+}}^{(a)}  + \mathcal{M}_{T_{cc}^{+}}^{(b)}, \nonumber  \\
\mathcal{M}_{T_{cc} \pi \rightarrow D^{*} D^{*}} &= & \mathcal{M}_{T_{cc}^{+}}^{(c)}  + \mathcal{M}_{T_{cc}^{+}}^{(d)}, \nonumber  \\
\mathcal{M}_{T_{cc} \rho \rightarrow D^{*} D} &= & \mathcal{M}_{T_{cc}^{+}}^{(e)}  + \mathcal{M}_{T_{cc}^{+}}^{(f)}, \nonumber \\
\mathcal{M}_{T_{cc} \pi \rightarrow DD^{*}} &= & \mathcal{M}_{T_{cc}^{+}}^{(g)}, \nonumber \\
\mathcal{M}_{T_{cc} \rho \rightarrow DD} &= & \mathcal{M}_{T_{cc}^{+}}^{(h)}  + \mathcal{M}_{T_{cc}^{+}}^{(i)}, \nonumber \\
\mathcal{M}_{T_{cc} \rho \rightarrow D^{*} D^{*}} & = & \mathcal{M}_{T_{cc}^{+}}^{(j)}  + \mathcal{M}_{T_{cc}^{+}}^{(k)}, 
      \label{ampl1}
\end{eqnarray}
where the explicit expressions are 
\begin{eqnarray}
\nonumber \mathcal{M}_{ T_{cc}^{+}}^{(a)} &\equiv &  g_{T_{cc} D D^{*}} g_{\pi D D^{*}} \epsilon_{1}^{\mu} \frac{1}{t – m_{D^{*}}^{2}}  
\\
	& &\times 
	\left( -g_{\mu \nu} + \frac{ (p_{1} – p_{3})_{\mu} (p_{1} – p_{3})_{\nu}}{m_{D^{*}}^{2}} \right) (p_{2} + p_{4})^{\nu} ,
\nonumber \\
\mathcal{M}_{ T_{cc}^{+}}^{(b)} &\equiv & -g_{T_{cc} D D^{*}} g_{\pi D D^{*}} \epsilon_{1}^{\mu} \frac{1}{u – m_{D^{*}}^{2}} 
\nonumber \\
	& &\times 
	\left( -g_{\mu \nu} + \frac{ (p_{1} – p_{4})_{\mu} (p_{1} – p_{4})_{\nu}}{m_{D^{*}}^{2}} \right) (p_{2} + p_{3})^{\nu} , \nonumber \\
	\mathcal{M}_{T_{cc}^{+}}^{(c)} &\equiv & -g_{T_{cc} DD^{*}} g_{\pi D D^{*}} \epsilon_{1}^{\mu} \epsilon_{4}^{\nu} \epsilon_{3 \mu}^{*}  \frac{1}{t – m_{D}^{2}} (2p_2 - p_4)_{\nu},
\nonumber \\
	\mathcal{M}_{T_{cc}^{+}}^{(d)} &\equiv & -g_{T_{cc} DD^{*}} g_{\pi D D^{*}} \epsilon_{1}^{\mu} \epsilon_{3 }^{\nu} \epsilon_{4\mu}^{*} \frac{1}{u – m_{D}^{2}} (2p_2 - p_3)_{\nu},
\nonumber \\
	\mathcal{M}_{T_{cc}^{+}}^{(e)} &\equiv & -g_{T_{cc} DD^{*}} g_{\rho DD} \epsilon_{1}^{\mu} \epsilon_{2}^{\nu} \epsilon_{3 \mu}^{*} \frac{1}{t – m_{D}^{2}} (2p_{4} – p_{2})_{\nu}, 
\nonumber \\
\nonumber \mathcal{M}_{T_{cc}^{+}}^{ (f) } &\equiv & -g_{T_{cc} DD^{*}} g_{\rho D^{*} D^{*}} \epsilon_{1}^{\mu} \epsilon _{2}^{\alpha} \epsilon _{3}^{* \beta} \frac{1}{u – m_{D^{*}}^{2}} 
\\
\nonumber & & \times 
\left( -g_{\mu \nu} + \frac{ (p_{1} – p_{4})_{\mu} (p_{1} – p_{4})_{\nu} }{ m_{D^{*}}^{2} } \right) 
\nonumber \\
	& &\times 
	((2p_{3} – p_{2})_{\alpha} g_{\beta}^{\nu} – (p_{3} + p_{2})^{\nu} g_{\alpha \beta} + (2p_{2} – p_{3})_{\beta} g_{\alpha}^{\nu}),
\nonumber \\
   \label{ampl2}
\end{eqnarray}
and 
\begin{eqnarray}
	\mathcal{M}_{T_{cc}^{+}}^{(g)} &\equiv & g_{\pi D^{*} D^{*}} g_{T_{cc} D D^{*}}  \epsilon_{1}^{\mu} \epsilon_{4}^{\alpha} \frac{1}{t – m_{D^{*}}^{2}} \varepsilon_{\mu \nu \alpha \beta} p_{2}^{\nu} p_{4}^{\beta},
\nonumber \\
\nonumber \mathcal{M}_{T_{cc}^{+}}^{(h)} &\equiv & g_{\rho D D^{*}} g_{T_{cc}DD^{*}} \epsilon_{1}^{\mu} \epsilon_{2 \alpha}^{*} \frac{1}{t – m_{D^{*}}^{2}}
 \\ 
	& & \times 
	\left( -g_{\mu \nu} + \frac{ (p_{1} – p_{3})_{\mu} (p_{1} – p_{3})_{\nu}  }{ m_{D^{*}}^{2}} \right) \varepsilon^{\nu \alpha \beta \gamma} p_{2 \beta} p_{4 \gamma},
\nonumber \\
\nonumber \mathcal{M}_{T_{cc}^{+}}^{(i)} &\equiv &  -g_{\rho D D^{*}} g_{T_{cc}DD^{*}} \epsilon_{1}^{\mu} \epsilon_{2 \alpha}^{*} \frac{1}{u – m_{D^{*}}^{2}} 
\nonumber \\
	& & \times 
	\left( -g_{\mu \nu} + \frac{ (p_{1} – p_{4})_{\mu} (p_{1} – p_{4})_{\nu}  }{ m_{D^{*}}^{2}} \right) \varepsilon^{\nu \alpha \beta \gamma} p_{2 \beta} p_{3 \gamma},
\nonumber \\
\mathcal{M}_{T_{cc}^{+}}^{(j)} &\equiv & -g_{\rho D D^{*}} g_{T_{cc}DD^{*}} \epsilon_{1}^{\mu} \epsilon_{3 \mu}^{*} \epsilon_{2}^{\nu} \epsilon_{4}^{\alpha} \frac{1}{t – m_{D}^{2}} \varepsilon_{\nu \alpha \beta \gamma} p_{2}^{\beta} p_{4}^{\gamma}, 
\nonumber \\
	\mathcal{M}_{T_{cc}^{+}}^{(k)} &\equiv & g_{\rho D D^{*}} g_{T_{cc}DD^{*}} \epsilon_{1}^{\mu} \epsilon_{4 \mu}^{*} \epsilon_{2}^{\nu} \epsilon_{3}^{\alpha} \frac{1}{u – m_{D}^{2}} \varepsilon_{\nu \alpha \beta \gamma} p_{2}^{\beta} p_{3}^{\gamma}. 
      \label{ampl3}
\end{eqnarray}
In the above equations, $p_1$ and $p_2$ are the momenta of initial state 
particles, while $p_3$ and $p_4$ are those of final state particles;         
$\epsilon_i ^{ \mu} \equiv \epsilon^{ \mu} (p_i)$ is the polarization vector 
related to the  respective vector particle $i$; $t$ and $u$ are the 
Mandelstam variables, which together with the $s$-variable they are
defined as: $s = (p_1 +p_2)^2, t = (p_1 - p_3)^2,$ and $u = (p_1-p_4)^2$.

We define the total isospin-spin-averaged cross section in the center of mass
(CM) frame for the processes in Eq. (\ref{ampl1}) as
\begin{eqnarray}
  \sigma_{a b \rightarrow c d}  
= \frac{1}{64 \pi^2 s }  \frac{|\vec{p}_{cd}|}{|\vec{p}_{ab}|}  \int d \Omega 
\overline{\sum_{S, I}} 
|\mathcal{M}_{a b \rightarrow c d} |^2 ,
\label{eq:CrossSection}
\end{eqnarray}
where $ a b \rightarrow c d $ designates the reaction according to     
Eq.~(\ref{ampl1}); $\sqrt{s}$ denotes the CM energy;  $|\vec{p}_{ab}|$   
and $|\vec{p}_{cd}|$ are  the three-momenta of initial and final particles   
in the CM frame, respectively; $d \Omega = d\phi d(\cos{(\theta)})$ is the 
solid angle measure; the symbol $\overline{\sum_{S,I}}$ stands for the sum 
over the spins and isospins of the particles, weighted by the isospin and 
spin degeneracy factors of the two particles forming the initial state,
i.e. \cite{XProd1} 
\begin{eqnarray}
\overline{\sum_{S,I}}|\mathcal{M}_{a b \rightarrow c d}|^2 & \to & 
\frac{1}{g_{a}}
 \frac{1}{g_{b}} \sum_{S,I}|\mathcal{M}_{a b \rightarrow c d}|^2, 
\label{eq:DegeneracyFactors}
\end{eqnarray}
with $g_{a}=(2I_{a}+1)(2S_{a}+1)$ and $g_{b}= (2I_{b}+1)(2S_{b}+1)$ are 
the degeneracy factors  of the particles in the initial state. In the
present analysis we do not consider isospin violation. 

The cross sections of the inverse processes, in which  $T_{cc}^{+}$
is produced,  can be calculated using the detailed balance relation, i.e.
\begin{eqnarray}
 g_{a}g_{b} |\vec{p}_{ab}|^2 \sigma_{ a b  \rightarrow c d } (s) = 
 g_{c} g_{d}  |\vec{p}_{cd}|^2   \sigma_{c d   \rightarrow a b } (s).
\label{detailedbalanceeq}
\end{eqnarray}

Finally, in the implementation of the numerical calculations another    
ingredient should be brought into play.To avoid the artificial increase 
of the amplitudes with the CM energy, we must include a form factor for
each vertex present in a given diagram. This is the subject of the next
section. 

\section{Form Factors and Couplings }
\label{FormFactors}

The theory of form factors is the theory of three-point (or four-point)
correlation functions associated with a generic vertex of three mesons $M_1$,
$M_2$ and $M_3$. A three-point correlation function, which depends on the
external 4-momenta $p$ and $p'$ is given by:
\begin{eqnarray}
  \Gamma(p,p') & = & \int d^4 x \,  d^4 y \, e^{i p'. x} \, e^{-i(p - p').y}
  \nonumber \\
& & \times \langle 0| T \{ j_3(x) j_2^{\dagger}(y) j_1^{\dagger}(0)\}|0\rangle
\label{3pcorrf}
\end{eqnarray}
where the $T$ is the time-ordered product and the current $j_i$ represents
states  with
the quantum numbers of the
meson $i$. This correlation function is evaluated in two ways. In the first one,
we consider that the currents are composed by quarks and write them in terms of
their flavor and color content with the correct quantum numbers. This is the QCD
description of the correlator, also known as OPE (Operator Product Expansion)
description.  In the second way, we write the correlation
function in terms of matrix elements of hadronic states which can be extracted
from experiment, or calculated with lattice QCD or estimated with effective
Lagrangians. In this second approach we never talk about quarks and use
all the available experimental information concerning the masses and decay
properties of the relevant mesons. This is the hadronic description of the
correlator, also known as Phenomenological description.  After performing the
two evaluations of the correlation function
separately, we identify one description with the other, obtaining an equation.
In this equation the form factor, i.e. the function
$g_{M_1 \,  M_2 \, M_3} (p,p')$, is the unknown, which is determined in terms
of the QCD parameters (quark masses and couplings) and also in terms of the
meson masses and decay constants.   This  procedure can be implemented in    
lattice QCD. In its analytic (and approximated) version it is called QCD sum
rules (QCDSR)\cite{svz,rry}.  In Ref.~\cite{bcnn11} all the form
factors required for the present calculation were computed in QCDSR, except for
the $T_{cc}-D-D^*$  form factor, which will be calculated in the next subsection. 

\subsection{The $T_{cc}-D-D^*$  Form Factor}

In this section  we use QCDSR 
to study the $T_{cc}$  form factor in the vertex  $T^+_{cc} D^0{D}^{*+}$,   
considering $T_{cc}$ as a four-quark state. The form factor in the         
vertex  $T^+_{cc} D^+{D}^{*0}$ is, of course,  the same. Assuming that the
quantum numbers
of the $T_{cc}$ are $J^{P}=1^{+}$, the interpolating field for $T_{cc}^+$
is given by \cite{Navarra:2007yw}:
\beq
j_\mu=i(c_a^TC\gamma_\mu c_b)
(\bar{u}_a\gamma_5C\bar{d}_b^T)\;,
\label{field}
\enq
where $a,~b$ are color indices, and $C$ is the charge conjugation matrix.

The QCDSR calculation of the vertex  $T_{cc}^+ D^0{D}^{*+}$ is based on
the three-point function given by:
\beq
\Pi_{\alpha \mu}(p,\pli,q)=\int d^4x~ d^4y ~e^{i\pli.x}~e^{iq.y}~
\Pi_{\alpha \mu}(x,y),
\lb{3po}
\enq
with
\beq
\Pi_{\alpha \mu}(x,y)=\lag 0 |T[j_\alpha^{D^*}(x)j_5^{D}(y)
j_\mu^\dagger(0)]|0\rag,
\enq
where $p=\pli+q$ and the interpolating fields for $D^0$ and $D^{*+}$
are given by: 
\beq
j_{5}^{D}=i\bar{u}_a\gamma_5 c_a,\mbox{ and }
j_{\alpha }^{D^*}=\bar{d}_a\gamma_\alpha  c_a.
\lb{D}
\enq
In order to evaluate the phenomenological side of the sum rule we
insert intermediate states for $T_{cc}$, $D$ and $D^*$ into Eq.(\ref{3po}).
We get:
\beqa
&&\Pi_{\alpha \mu}^{(phen)} (p,\pli,q)={-i\lambda_{T_{cc}} m_{D^*}f_{D^*}f_{D}
m^2_{D}~g_{T_{cc}DD^*}(q^2)
\over m_c(p^2-m_{T_{cc}}^2)({\pli}^2-m_{D^*}^2)(q^2-m_D^2)}
\nnb\\
&\times&\left(-g_{\alpha \lambda}+{\pli_\alpha \pli_\lambda\over m_{D^*}^2}\right)
\left(-g_\mu^\lambda+{p_\mu p^\lambda\over m_{T_{cc}}^2}\right)
+\cdots\;,
\lb{phen3}
\enqa 
where the dots stand for the contribution of all possible excited states.
The form factor, $g_{T_{cc}DD^*}(q^2)$, is defined as the generalization
of the on-mass-shell matrix element, $\lag D^* \,  D \,| \, T_{cc}\rag$,
for an off-shell $D$ meson:
\beq
\lag D^*(\pli) D(q)|T_{cc}(p)\rag=g_{T_{cc}DD^*}(q^2)
\varepsilon^*_\lambda(\pli)\varepsilon^\lambda(p),
\label{coup}
\enq
where
$\varepsilon_\mu(p),~\varepsilon_\alpha(\pli)$ are  the polarization
vectors for $T_{cc}$ and $D^*$ mesons  respectively.
In deriving  Eq.~(\ref{phen3}) we have used the definitions:
\beqa
\lag 0 | j_\alpha^{D^*}|D^*(\pli)\rag &=&m_{D^*} f_{D^*}\varepsilon_\alpha(\pli),
\nnb\\
\lag 0 | j_{5}^{D}|D(q)\rag &=& {f_{D}m^2_{D}\over m_c},
\nnb\\
\lag T_{cc}(p) | j_\mu|0\rag &=&\lambda_{T_{cc}}\varepsilon_\mu^*(p).
\lb{fp}
\enqa
The definition of the above matrix
elements  is a matter of convention. The definition in the
second line  is the usual one, adopted for example, in Eq. (5) of
Ref.~\cite{kho95}.   However, it is also possible to  include the quark
mass in the definition of the current, as it was done in Eq. (2.4) of the
recent paper \cite{kho21}. 

As discussed  refs.~\cite{Dias:2013xfa}, large partial
decay widths are
expected when the coupling constant is obtained from QCDSR in the case of
multiquark states which  contain
the same number of valence quarks as the number of valence quarks in
the final state. This happens  because, although the initial current,
Eq.~(\ref{field}), has a non-trivial color structure, it can be rewritten as
a sum of molecular type currents with trivial color configuration through a
Fierz transformation. To avoid this problem we follow
Ref.~\cite{Dias:2013xfa}, and consider in the OPE side of the sum rule only     
the diagrams with non-trivial color structure, which are called color-connected
(CC) diagrams. Isolating the $\pli_\mu\pli_\alpha$ structure in Eq.~(\ref{phen3})
we have \cite{Dias:2013xfa}: 
\beqa
\Pi^{(OPE)}&=& {-i m_c\mix \over48\sqrt{2}\pi^2}\left[{1\over m_c^2-q^2}
\int_0^1 d\alpha{\alpha(2+\alpha)\over m_c^2-(1-\alpha){\pli}^2}\right.
\nnb\\
&-&\left.{1\over m_c^2-{\pli}^2}
\int_0^1 d\alpha{\alpha(2+\alpha)\over m_c^2-(1-\alpha)q^2}\right].
\label{ope3}
\enqa
where $\mix$ is the mixed quark-gluon condensate. Equating
$\Pi_{\alpha \mu}^{(phen)} (p,\pli,q)$ to
$\Pi^{(OPE)} (p,\pli,q)$, using the euclidean four-momenta ($P^2 = -p^2$, 
${P^\prime}^2 = - {p^\prime}^2$) and performing a single Borel transformation
on both momenta $P^2={P^\prime}^2\rightarrow M^2$, we get the sum rule:
\beqa
&&{1\over Q^2+m_D^2}\left[A\left(e^{-m_{D^*}^2/M^2}
  -e^{-m_{T_{cc}}^2/M^2}\right)+
B~e^{-s_0/M^2}\right]=
\nnb\\
&&{m_c\mix \over48\sqrt{2}\pi^2}\left[{1\over m_c^2+Q^2}
  \int_0^1 d\alpha{\alpha(2+\alpha)\over1-\alpha}~
  e^{- m_c^2\over \alpha(1-\alpha)M^2}\right] 
\nnb\\ 
&-&\left.e^{-m_c^2/M^2}
\int_0^1 d\alpha{\alpha(2+\alpha)\over m_c^2+(1-\alpha)Q^2}\right],
\label{sr3}
\enqa
where $Q^2 = -q^2$ is the euclidean four momentum of the off-shell $D$
meson, $s_0$ is the continuum threshold parameter for $T_{cc}$,
\beq
A={g_{T_{cc}DD^*}(Q^2)\, \lambda_{T_{cc}} \,  f_{D^*} \, f_{D}m_{D}^2
\over m_c \, m_{D^*} \, (m_{T_{cc}}^2-m_{D^*}^2)},
\label{e}
\enq
and $B$ is a parameter introduced to take into account single pole
contributions associated with pole-continuum transitions,
which are not suppressed  when only a single Borel transformation
is done in a three-point function sum rule \cite{io1}.
In the numerical analysis we use the following values for quark masses and QCD
condensates \cite{bcnn11}:
\beqa
\label{qcdparam}
&m_c(m_c)=(1.23\pm 0.05)\,\GeV,\nnb\\
&\qq = - (0.23\pm0.03)^3
\,\GeV^3,\nnb\\
&\lag\bar{q}g\si.Gq\rag=m_0^2\lag\bar{q}q\rag,\nnb\\
&m_0^2=
0.8\,\GeV^2.
\enqa
We use the experimental values for $ m_D$ and $m_{D^*}$
\cite{Zyla:2020zbs} and we take $f_{D}$ and $f_{D^*}$ from 
Ref.~\cite{bcnn11}:
\beqa
m_D=1.869~\GeV,\;\;f_{D}=(0.18\pm0.02)~\GeV,
\nnb\\
m_{D^*}=2.01~\GeV,\;\;f_{D^*}=(0.24\pm0.02)~\GeV. 
\enqa
The meson-current coupling,
$\lambda_{T_{cc}}$, defined in Eq.(\ref{fp}), can be determined from the
two-point  sum rule \cite{Navarra:2007yw}:
$\lambda_{T_{cc}}=(2.2\pm0.3)\times10^{-2}~\GeV^5$,
and we take the mass of the $T_{cc}$ from \cite{LHCb:2021auc}:
$m_{T_{cc}}=(3874.817 \pm0.061)$ MeV.
For the continuum threshold we use $s_0=(m_{T_{cc}}+\Delta s_0)^2$,
with $\Delta s_0= (0.5\pm0.1)~\GeV$. 
One can use Eq.~(\ref{sr3}) and its derivative with respect to $M^2$ to
eliminate $B$ from Eq.~(\ref{sr3}) and to isolate $ g_{T_{cc}DD^*}(Q^2)$.
A good Borel window is determined when the parameter to be
extracted from the sum rule is as much independent of the Borel mass
as possible. Analysing $g_{T_{cc}DD^*}(Q^2)$, as a function of
both $M^2$ and $Q^2$, we find a very good Borel stability 
in the region $2.2 \leq M^2 \leq 2.8$ GeV$^2$. Fixing $M^2=2.6$ GeV$^2$       
we can extract the $Q^2$ dependence of the $g_{T_{cc}DD^*}(Q^2)$ form factor.
Since the coupling constant is defined as the value of the form factor at the
meson pole: $Q^2 = - m^2_{D}$,  we need to extrapolate
the form factor for a region of $Q^2$ where
the QCDSR are not valid. This extrapolation
can be done by parametrizing the QCDSR results for
$g_{T_{cc}DD^*}(Q^2)$ with the help of an exponential form: 
\beq
g_{T_{cc}DD^*}(Q^2) = g_{T_{cc}DD^*} \, e^{-g (Q^2+m^2_D)},
\label{exp}
\enq
with $ g=0.076~\mbox{GeV}^{-2}$.
For other values of the Borel mass, in  the range
$2.2 \leq M^2 \leq 2.8$ GeV$^2$, the results are equivalent.
We get for the coupling constant:
\beq
g_{T_{cc}DD^*}=g_{T_{cc}DD^*}(-m^2_D)=(1.7 \pm 0.2)~~\mbox{GeV}.
\label{coupdd}
\enq

The uncertainty in the coupling constant  comes from
variations in $s_0$, $\lambda_{T_{cc}},~f_D,~f_{D^*}$, $\mix$  and $m_c$.

In order to evaluate the diagrams shown in Fig. \ref{DIAG1} we need the
form factors in the vertices $D \pi D^* $, $D \rho D$, $D^* \rho D^*$, 
$D^* \pi D^*$ and $D \rho D^*$.  All these form factors have been calculated
with QCDSR \cite{bcnn11} and could be parametrized with the following forms:
\beq
I) \,\,\,\, g_{M_1 M_2 M_3} = \frac{A}{Q^2 + B}
\label{FI}
\enq
and
\beq
II) \,\,\,\, g_{M_1 M_2 M_3} = A \, e^{-(Q^2/B)} 
\label{FII}
\enq
where $M_1$ is the off-shell meson in the vertex and $Q^2$ is its euclidean
four momentum. The parameters $A$ and $B$ are given in Table \ref{FFparam}. 

\begin{table}[h] 
        \begin{center} 
                \begin{tabular}{cccc} 
                        \hline 
                        \hline 
                        $M_1$ $M_2$ $M_3$ & Form & A    & B     \\ 
                        \hline 
                        $D \pi D^* $      & I    & 126  & 11.9  \\ 
                        \hline 
                        $D \rho D$        & I    & 37.5 & 12.1  \\ 
                        \hline  
                        $D^* \rho D^*$    & II   & 4.9  & 13.3  \\ 
                        \hline  
                        $D^* \pi D^*$     & II   & 4.8  & 6.8   \\ 
                        \hline  
                        $D \rho D^*$      & I    & 234  & 44    \\ 
                        \hline  
                        \hline 
                \end{tabular} 
                \caption{Parameters for the form factors in the
                  $M_1$ $M_2$ $M_3$ vertex. The meson $M_1$ is off-shell.} 
                \label{FFparam} 
        \end{center} 
\end{table}

\subsection{Empirical Formulas}

The QCDSR method used in the previous section has an important limitation.
It is restricted to multiquark systems in a compact configuration. As it can
be seen in (\ref{3pcorrf}) and (\ref{field}) all the quark fields in the
current are defined at the same space-time point. This is a good   
approximation for tetraquarks. If the multiquark system was
in a meson-meson molecular configuration, the distance between the quarks would
have to be included explicitly (through a path-ordered gauge connection).
This would increase considerably the difficulty of the calculation and, so far, 
has not yet been incorporated in the QCDSR method. For this reason (and also
for simplicity) several authors have used empirical form factors with simple
monopole, dipole, exponential or gaussian forms. Some of these form factors
rely on a molecular picture of the multiquark system. Along this line,  
the best we can do is to estimate the systematic uncertainty related to the
form factors. To this end we have tried several functional forms, which were
already used in the literature,  and varied the corresponding cutoff 
parameters. After several tests, we have decided to consider two extreme cases.
The first is the ``softest'' form factor given by the
Gaussian form~\cite{Abreu:2018mnc}:
\begin{eqnarray}
	F & = & \exp{\left(- \frac{ (q^2 - m_{ex} ^2 )^2 }{\Lambda ^4}\right)},
	  \label{formfactor}
\end{eqnarray}
where $q$ is the four-momentum of the exchanged particle of mass $m_{ex}$ 
for a vertex involving a $t$- or $u$-channel meson exchange. As discussed
in~\cite{Abreu:2018mnc}, this form factor corresponds to the limit of the     
form $	[ n \Lambda^4 /( n \Lambda ^4 + (p^2 - m_{ex} ^2 ) )^2]^n  $  when    
$n\rightarrow \infty$.  The cutoff $\Lambda$ is chosen to be in 
the range $m_{min} < \Lambda < m_{max}$, where $m_{min}$ ($m_{max}$)  is the
mass of the lightest (heaviest)  particle entering or exiting the vertices. 
The second  is the ``hardest'' form factor given by  the monopole-like
~\cite{Hong:2018mpk} expression:
\begin{eqnarray}
	F & = & \frac{\Lambda ^2}{\Lambda ^2 + \vec{q}^2},
	  \label{formfactor2}
\end{eqnarray}
where $\vec{q}$ is the momentum of the exchanged particle in a       
$t$- or $u$-channel in the center of mass frame. 
In  Ref.~\cite{Hong:2018mpk} the authors make use
of a monopole form factor with $\Lambda = 1.0$ GeV, while in
Ref.~\cite{ChoLee1} the same form factor with          
$\Lambda = 2.0$ GeV is employed in the analysis of the $X(3872)$ state. 
In what follows we will also use this functional form and vary the cut-off
in this range. 

Having specified the form factors we need to fix the coupling
constants.  The vertices shown in Fig.~\ref{DIAG1} involve ordinary mesons,
except 
for the $T_{cc} D D^*$ vertex, in which the coupling strength is sensitive to
the structure of the $T_{cc}$, being thus different for tetraquarks
and molecules. In the last subsection we have seen how to obtain
$g_{T_{cc} D D^*}$ for tetraquarks. For molecules, it was estimated in 
Ref.~\cite{Ling:2021bir} and it was found to be
$g_{T_{cc} D D^*} = 6.17 - 6.40  \GeV$.


\vspace{0.5cm}

\section{Results and Discussion}

\label{Results} 


The evaluation of the doubly-charmed state absorption and production cross sections  will be done in this Section with the isospin-averaged masses for the light and heavy mesons reported in Ref.~\cite{Zyla:2020zbs}: 
$m_{\pi} = 137.28 \MeV,$ $m_{\rho} = 775.38 \MeV, $                      
$ m_{\bar{D}} = 1867.24 \MeV $ and $ m_{\bar{D}^*} = 2008.56 \MeV; $ for 
the $T_{cc}^+$ we use $  m_{T_{cc}} = 3874.75 \MeV$
~\cite{LHCb:2021vvq,LHCb:2021auc}. 

\subsection{Empirical Formulas}

Here we employ the empirical form factors introduced in
preceding Section. 
The cut-off $\Lambda$ of these functional forms is  
chosen to be in the range $2.5-3.5 \GeV$ and $1.0-2.0 \GeV$ in the case
of the Gaussian and monopole form factors, respectively. In this kind of
calculation the cut-off is the most important source of uncertainty.
Because we use a
range of values (instead of a single number) for $\Lambda$, our
results will
be given with uncertainty bands. We note that the uncertainty associated with
the cut-off is much larger than the one related to the coupling constant,
which is not shown in the plots. In spite of these uncertainties we can
still draw conclusions from our calculations.

\begin{widetext}

\begin{figure}[!ht]
    \centering
           \includegraphics[{width=8.0cm}]{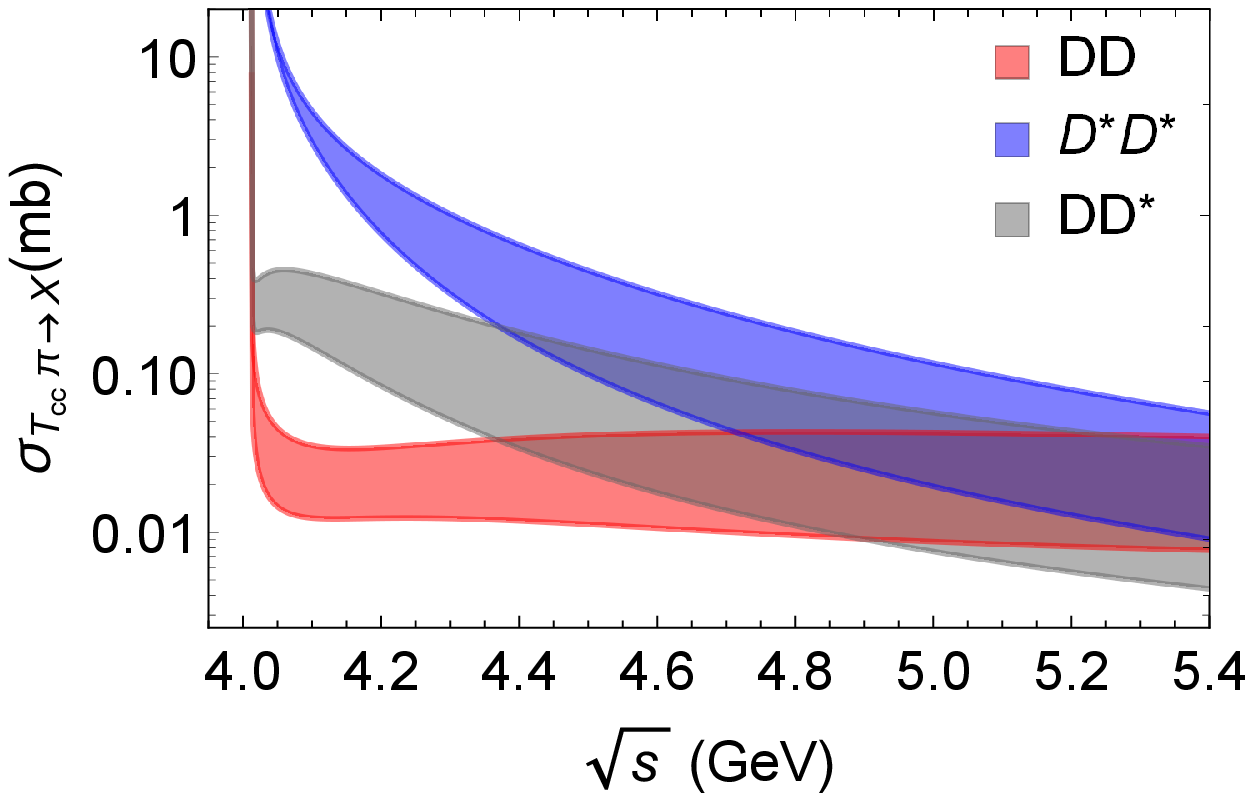}
           \includegraphics[{width=8.0cm}]{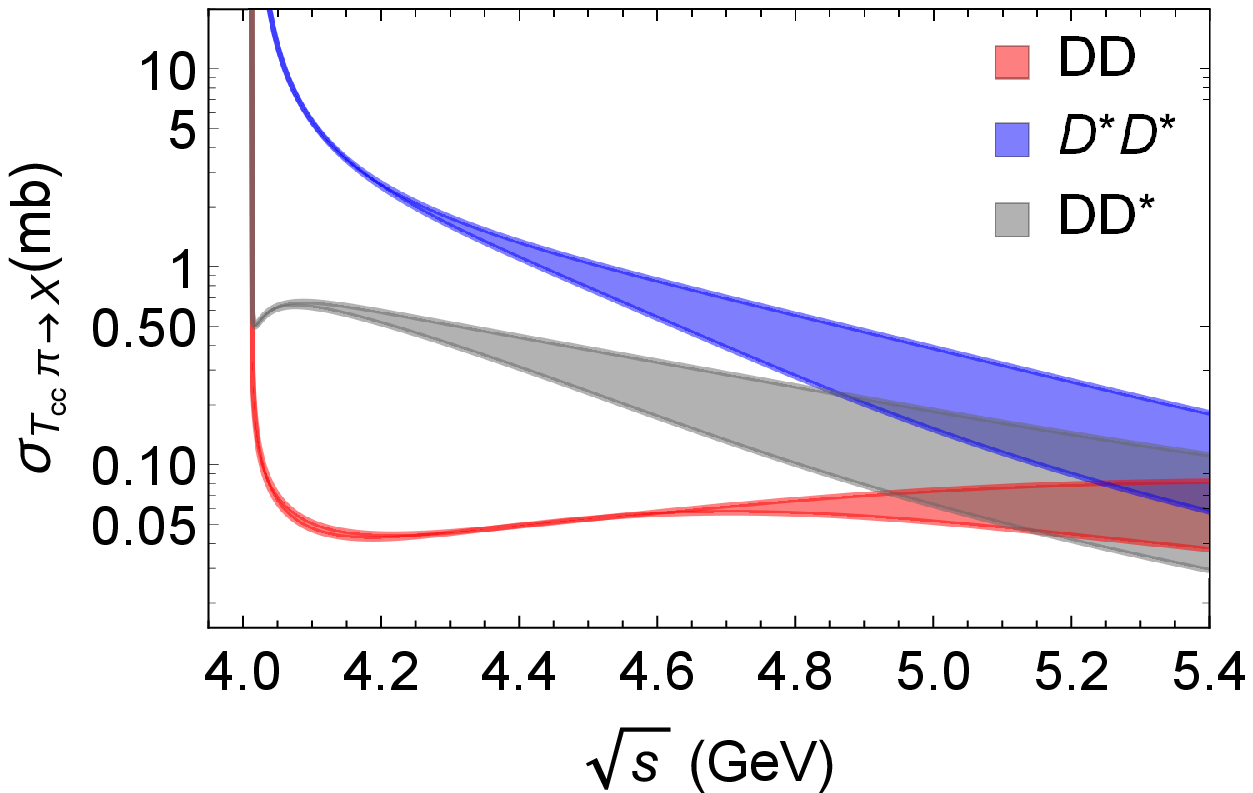} \\
           \includegraphics[{width=8.0cm}]{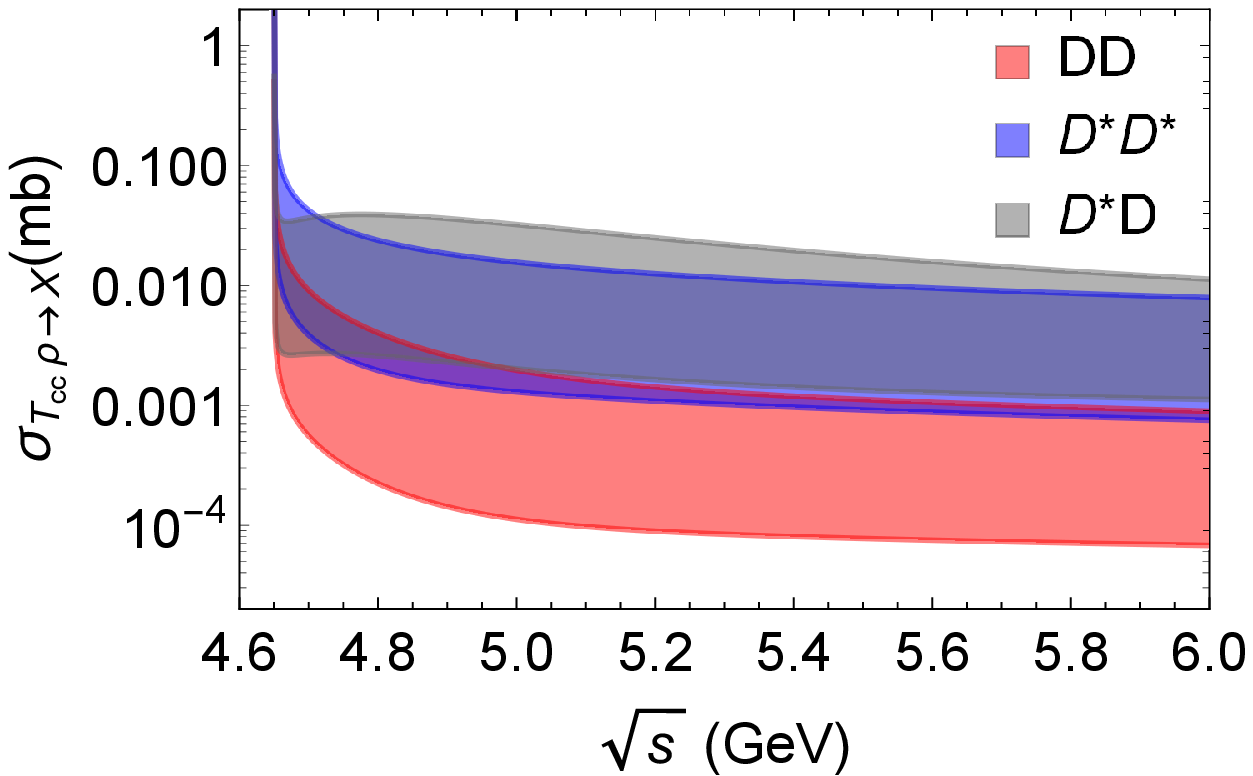}
           \includegraphics[{width=8.0cm}]{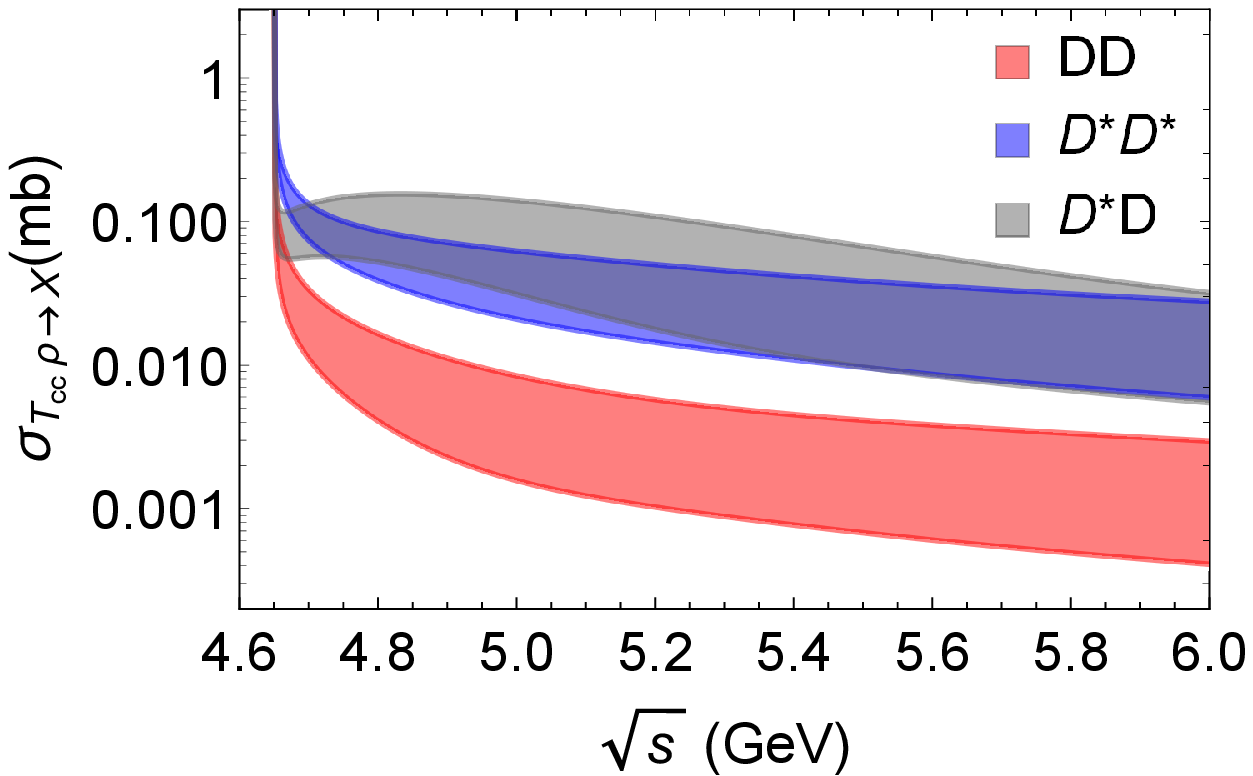}
 \caption{ Cross sections for the absorption processes
         $T_{cc}^+ \pi \rightarrow D^{(*)} D^{(*)} $ (top panels) and  
         $T_{cc}^+ \rho \rightarrow D^{(*)} D^{(*)} $ (bottom panels), 
as functions of the CM energy $\sqrt{s}$. Plots in left and right
panels: obtained by using monopole and Gaussian form factors, Eqs.
(\ref{formfactor}) and (\ref{formfactor2}) respectively. Upper and lower   
         limits of the band are obtained taking the upper
         and lower limits of the cutoff for each corresponding form factor
         ($1-2$ GeV and $2.5-3.5$ GeV for the monopole and Gaussian
         form factors, respectively). }          
    \label{Fig:CrSec-Abs-s}
\end{figure}

\end{widetext}

\begin{widetext}

\begin{figure}[!ht]
    \centering
           \includegraphics[{width=8.0cm}]{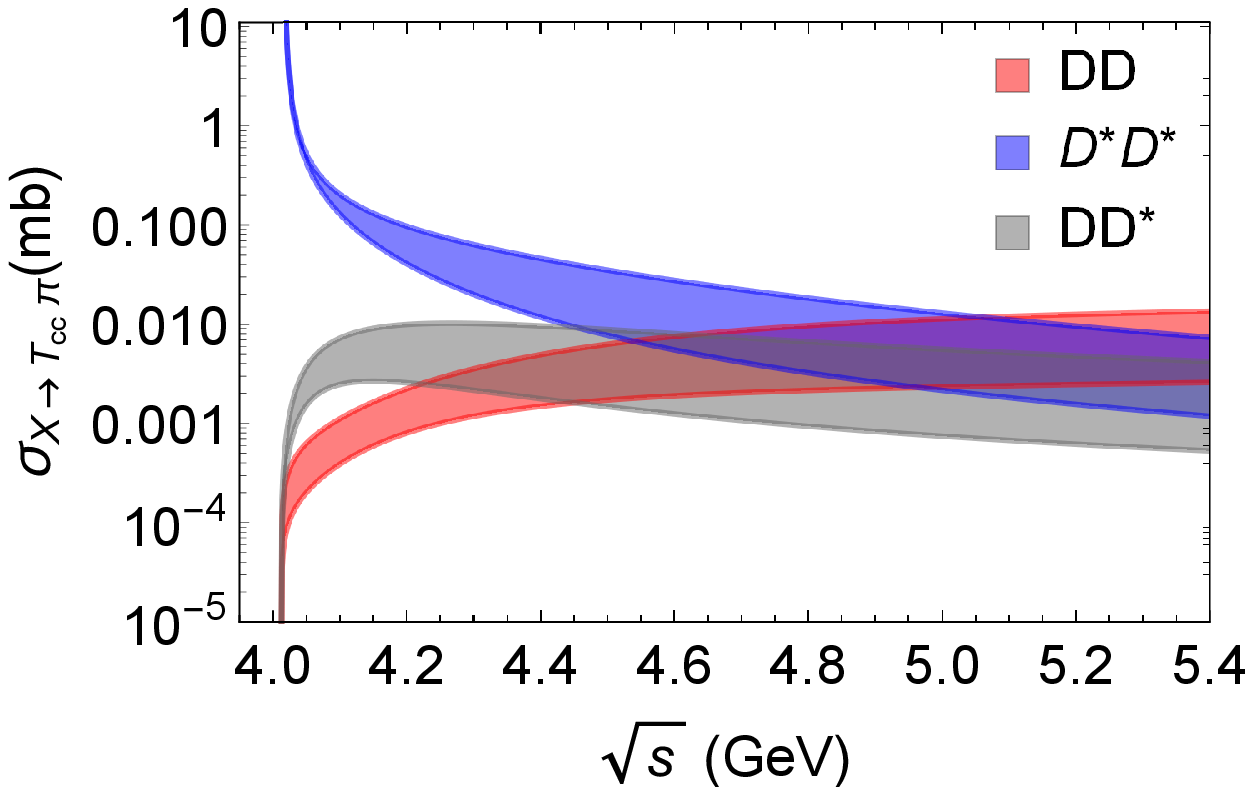}
           \includegraphics[{width=8.0cm}]{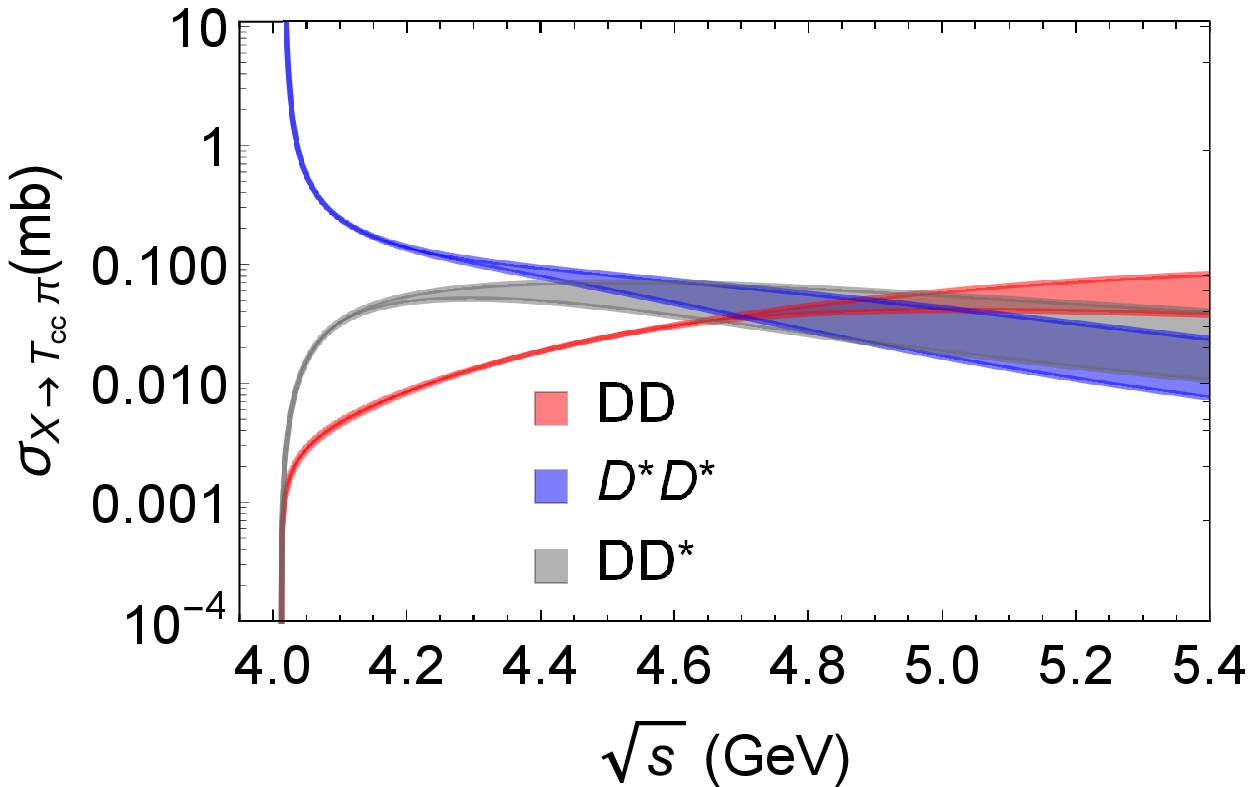} \\
           \includegraphics[{width=8.0cm}]{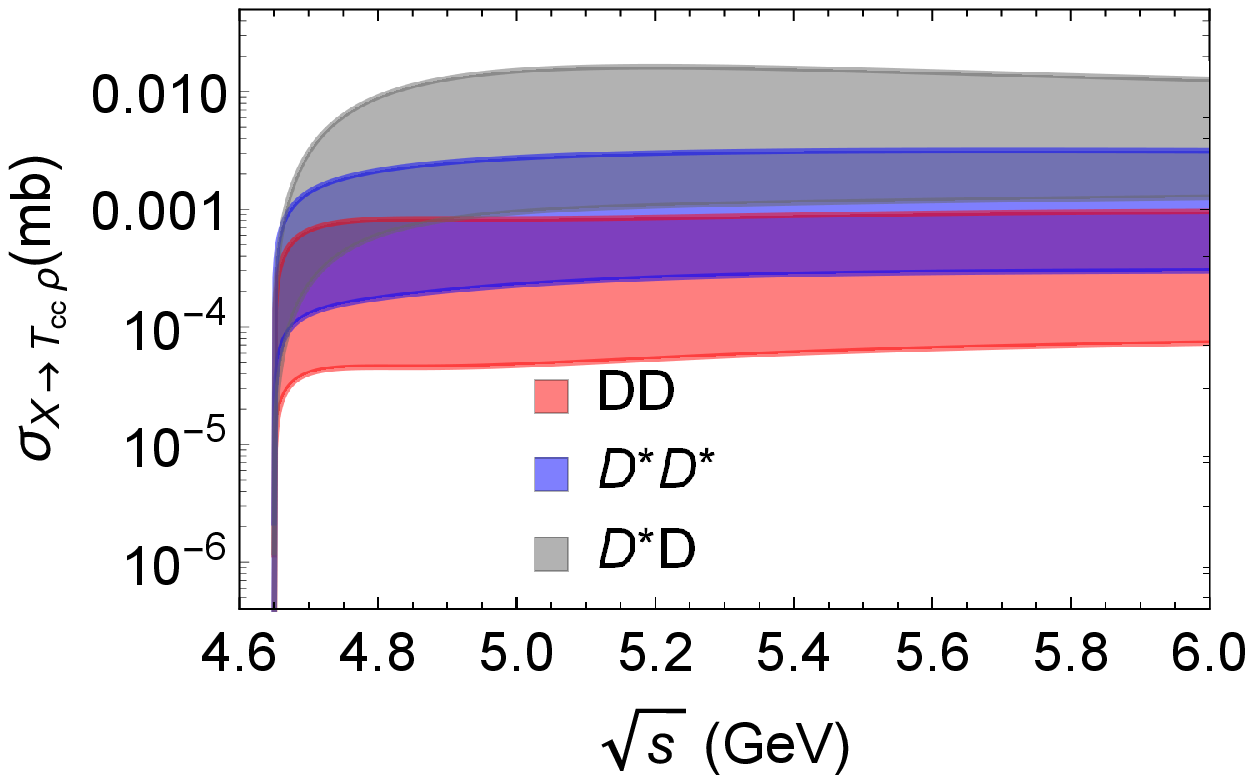}
           \includegraphics[{width=8.0cm}]{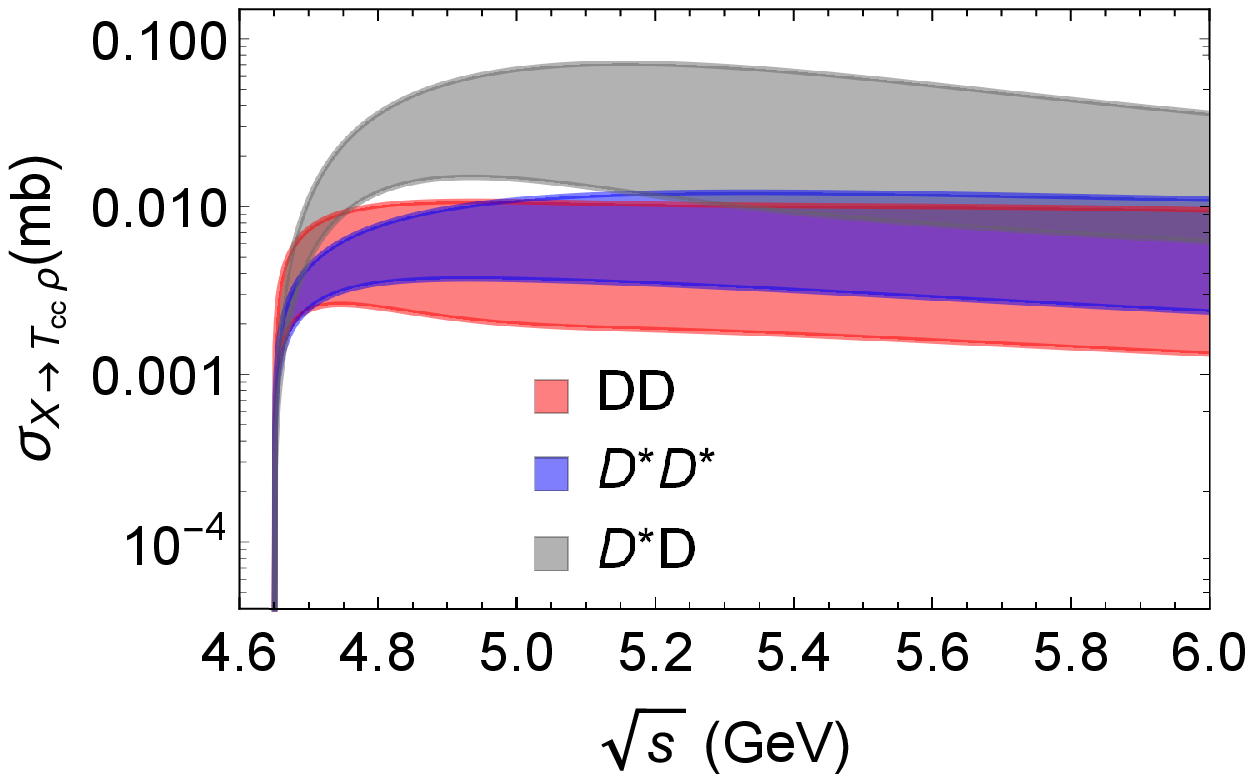}
       \caption{Cross sections as functions of the CM energy $\sqrt{s}$ 
         for the respective inverse (production) processes displayed in 
         Fig.~\ref{Fig:CrSec-Abs-s}, i.e.
         $D^{(*)} D^{(*)}\rightarrow T_{cc}^+ \pi  $ (top panels) and
         $D^{(*)} D^{(*)} \rightarrow T_{cc}^+ \rho $  (bottom panels),
         obtained via the detailed balance relation. Plots in left and
         right panels: obtained by using monopole and Gaussian form factors,
         Eqs. (\ref{formfactor}) and (\ref{formfactor}) respectively.
         Upper and lower   
         limits of the band are obtained taking the upper
         and lower limits of the cutoff for each corresponding form factor
         ($1-2$ GeV and $2.5-3.5$ GeV for the monopole and Gaussian form
         factors, respectively).
         }
    \label{Fig:CrSec-Prod-s}
\end{figure}
\end{widetext}

Plots of the cross sections as functions of the CM energy $\sqrt{s}$ for 
the $T_{cc}^+$-absorption by pion or $\rho$ mesons are shown in        
Fig.~\ref{Fig:CrSec-Abs-s}, using monopole and Gaussian form factors defined
respectively in Eqs. (\ref{formfactor}) and (\ref{formfactor2}).  Upper and  
lower limits of the bands are obtained taking the upper  and lower limits of 
the cutoff for each corresponding form factor, i.e.  $1-2$ GeV and $2.5-3.5$
GeV for the monopole and Gaussian form factors,  respectively.

Since all these absorption cross sections are exothermic (except the    
one for $T_{cc} \pi \rightarrow D^{*} D^{*}$), they become very large at
the threshold. We note that in the region close 
to the threshold the corresponding curves obtained with and without 
form factors are almost indistinguishable (we do not display another Figure
proving this for the sake of conciseness).   

The results suggest that within the range
$ 4.05 \leq \sqrt{s} \leq 4.5 \GeV$,
$\sigma _{T_{cc} \pi \rightarrow X}\, (X = D D, D D^{*},D^{*} D^{*})$  
have distinct magnitudes, being of order of $ \sim 10^{-3} -  1 \mb$.     
Close to the threshold, $\sigma _{T_{cc} \pi \rightarrow DD}$ is suppressed 
with respect to the other processes, at least by one order of magnitude.
However, at higher CM energies ($ 5 \leq \sqrt{s} \leq 5.5 \GeV$), all the 
processes have closer magnitudes.

Looking now at the absorption processes by a $\rho $ meson, which occur at 
higher thresholds, the $\sigma _{T_{cc} \rho \rightarrow D^* D, D^* D^*}$ are the
biggest at moderate energies, while the cross section for  $D D $ final      
states is the smallest. When we compare the different dissociation processes 
at a given CM energy, for instance $\sqrt{s} = 5 \GeV$,  we observe that
$\sigma _{T_{cc} \pi \rightarrow X}$ are greater than the respective
$\sigma _{T_{cc} \rho \rightarrow X}$  by about one order of magnitude. 
These findings allow to quantitatively estimate how big is the contribution
coming from the doubly-charmed state absorption by a pion with respect to
the other reactions.

Now let us move on to  the $T_{cc}^+$-production cross sections, which are
shown in Fig.~\ref{Fig:CrSec-Prod-s} as functions of the CM energy, again
using both monopole and Gaussian form factors. 
Except for $\sigma _{D^* D^* \rightarrow  T_{cc} \pi}$,  they are endothermic, 
with magnitudes of the order $ \sim  10^{-3} -  10^{-1} \mb$ in the case of
Gaussian form factor,  within the 
range of CM energies considered. For the monopole, the magnitudes are even
smaller. In general, the comparison with the outputs 
reported in Fig.~\ref{Fig:CrSec-Abs-s} reveals that both $T_{cc} \pi $ and 
$T_{cc} \rho $ absorption cross sections are greater than the respective  
production ones.  

%
%

\subsection{Form factors from QCDSR}


Now we employ the form factors given by Eqs.~(\ref{exp}), (\ref{FI}) and (\ref{FII}), obtained within the QCDSR approach. The bands in the figures 
express the uncertainty in the coupling constant $g_{T_{cc}DD^*}$ shown in
Eq.~(\ref{coupdd}).

\begin{figure}[!ht]
    \centering
           \includegraphics[{width=1.0\linewidth}]{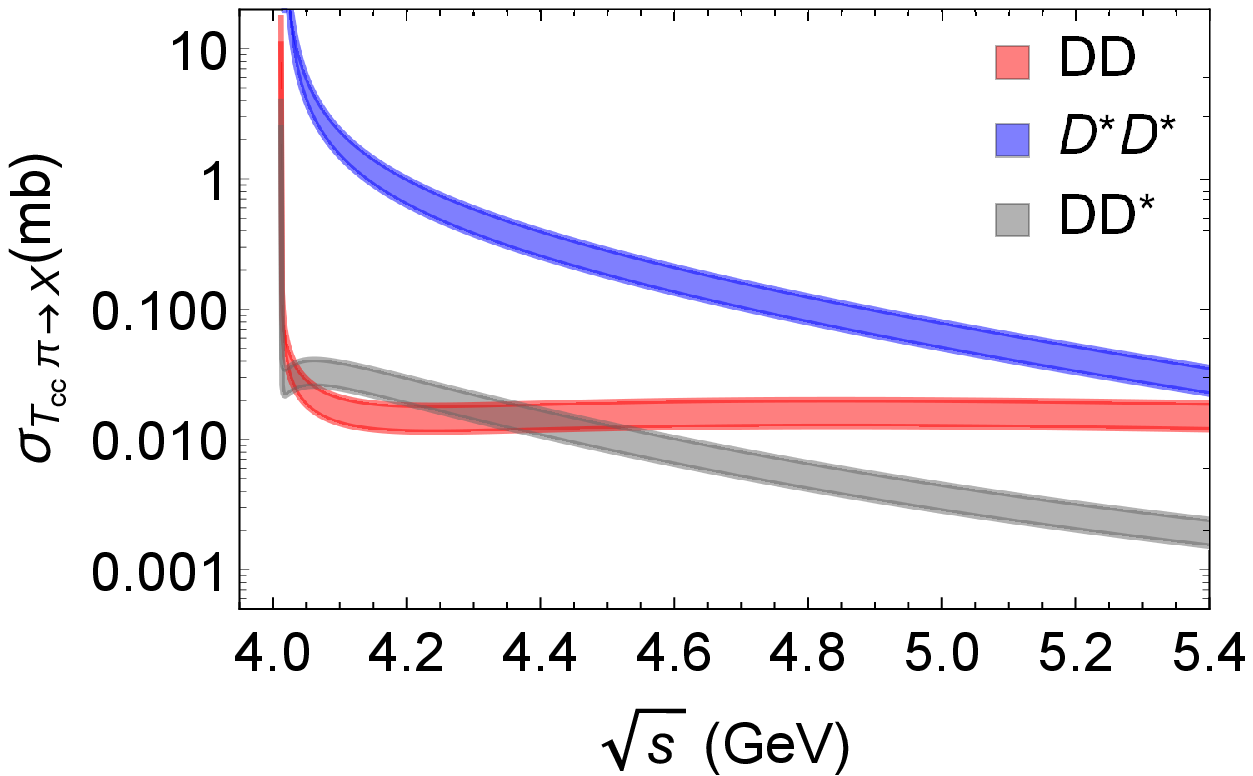} \\
           \includegraphics[{width=1.0\linewidth}]{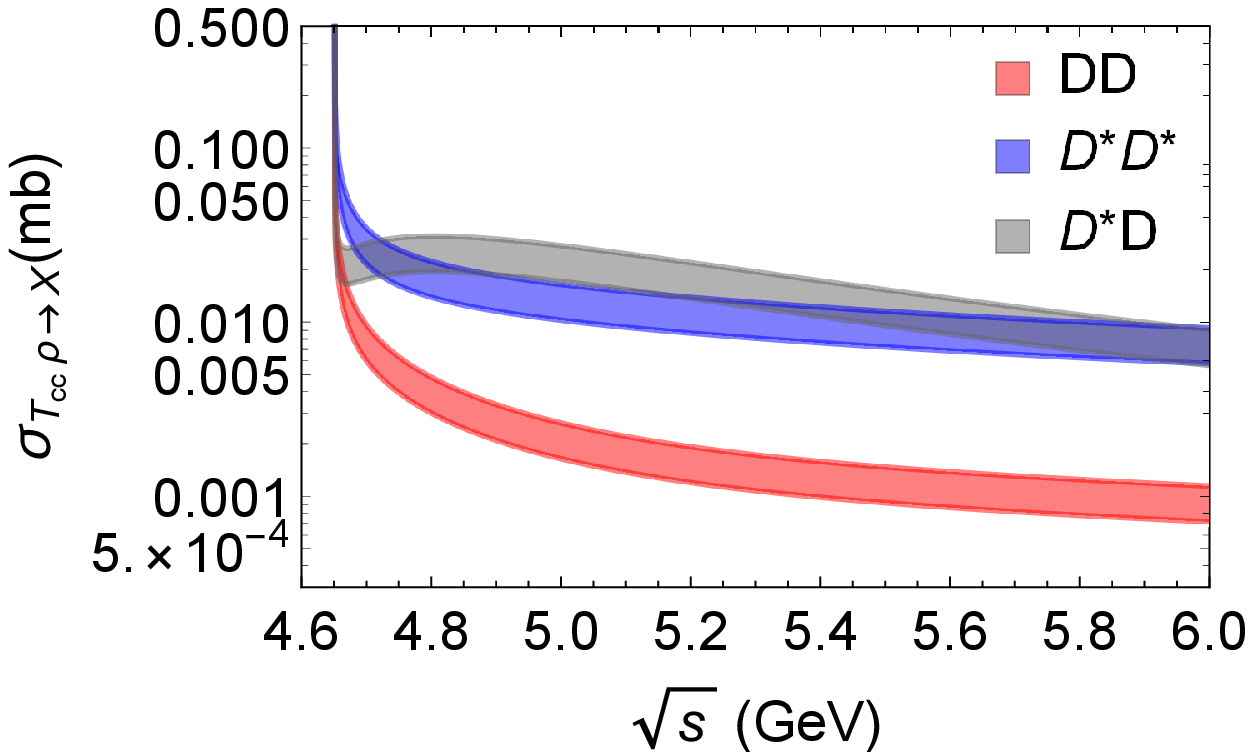}
 \caption{Cross sections for the absorption processes
         $T_{cc}^+ \pi \rightarrow D^{(*)} D^{(*)} $ (top panel) and  
         $T_{cc}^+ \rho \rightarrow D^{(*)} D^{(*)} $ (bottom panel), 
   as functions of the CM energy $\sqrt{s}$, taking the form factors obtained
   within the QCDSR approach.}          
    \label{Fig:CrSec-Abs-QCDSR}
\end{figure}

\begin{figure}[!ht]
    \centering
           \includegraphics[{width=1.0\linewidth}]{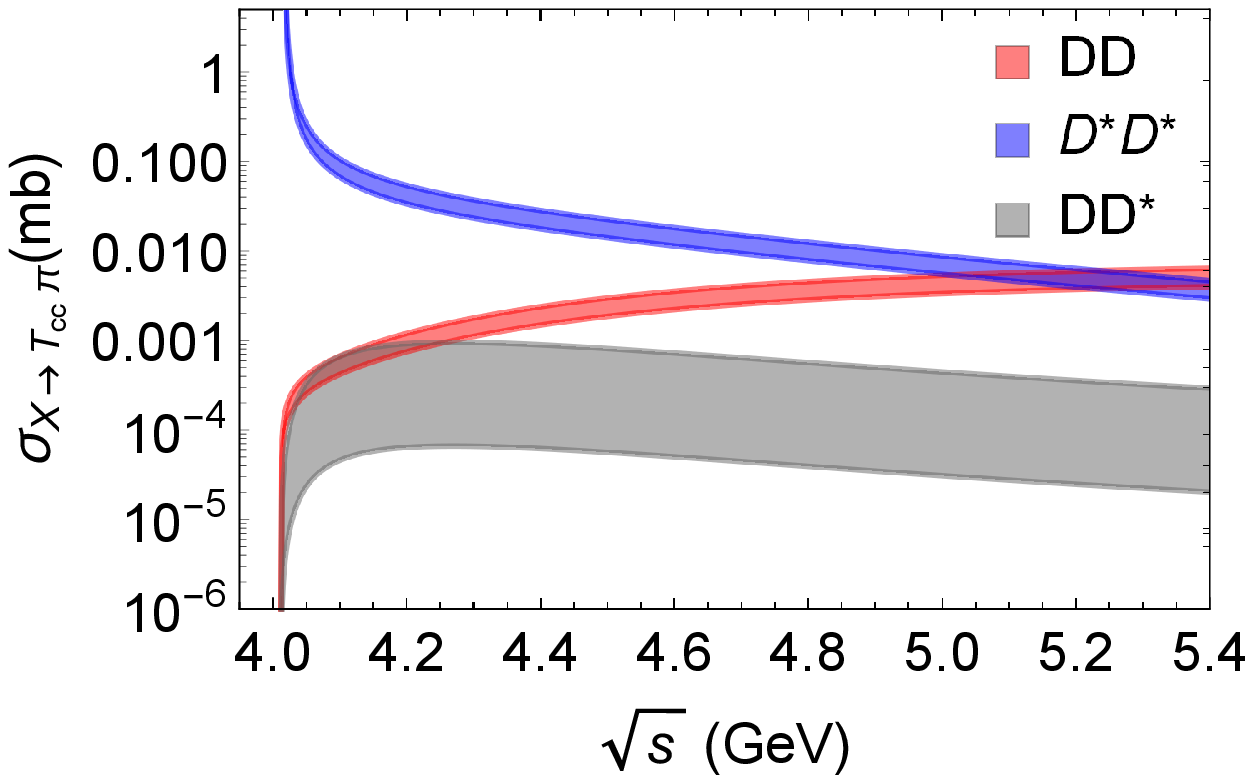}\\
           \includegraphics[{width=1.0\linewidth}]{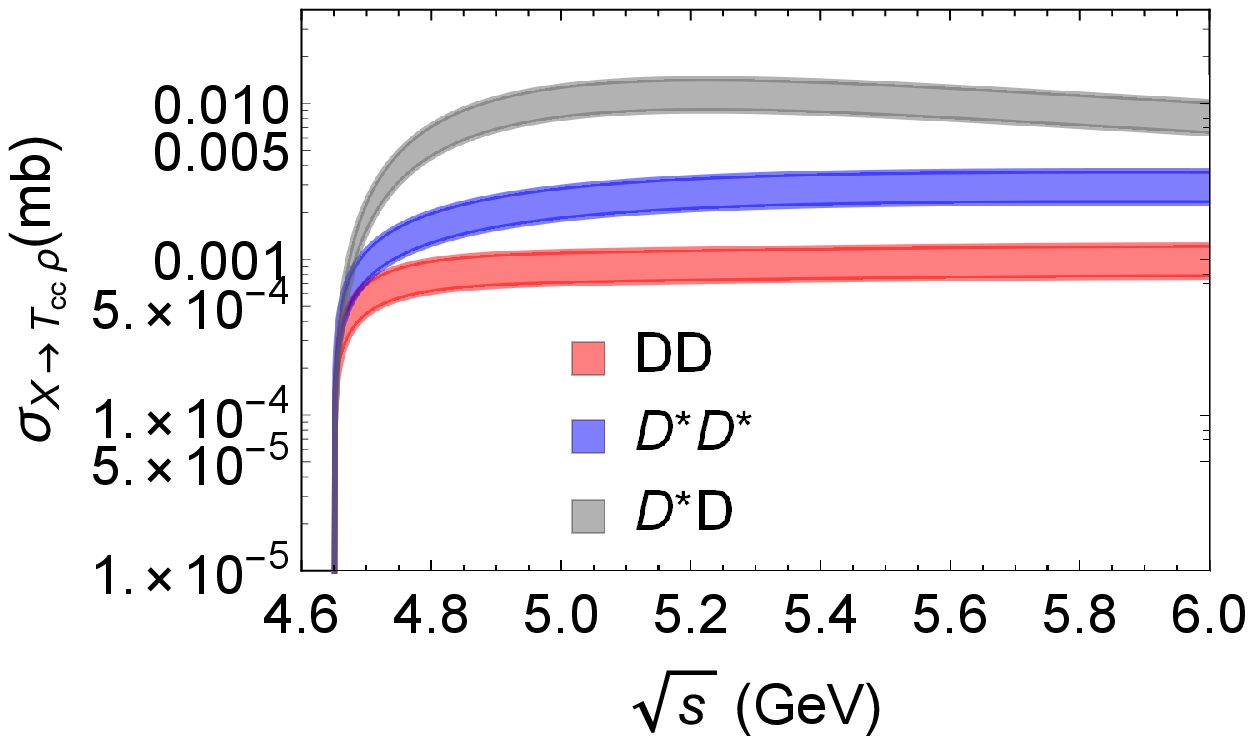}
       \caption{Cross sections as functions of the CM energy $\sqrt{s}$ 
         for the production processes
         $D^{(*)} D^{(*)}\rightarrow T_{cc}^+ \pi  $ (top panel) and   
         $D^{(*)} D^{(*)} \rightarrow T_{cc}^+ \rho $  (bottom panel),
         taking the form factors obtained within the QCDSR approach.).
         }
    \label{Fig:CrSec-Prod-QCDSR}
\end{figure}
 
Figs.~\ref{Fig:CrSec-Abs-QCDSR} and ~\ref{Fig:CrSec-Prod-QCDSR} show the plots 
of the cross sections as functions of the CM energy $\sqrt{s}$ for 
the $T_{cc}^+$-absorption and production by pion or $\rho$ mesons. 
Some qualitative features found in the preceding subsection    
(with the empirical form factors) are reproduced. Close to the threshold,
cross section $\sigma _{T_{cc} \pi \rightarrow D^*D^*}$
is greater than the cross sections for other processes.
Also, at $\sqrt{s} = 5 \GeV$, most of  
cross sections $\sigma _{T_{cc} \pi \rightarrow X}$ are greater than the      
respective $\sigma _{T_{cc} \rho \rightarrow X}$, although this difference is
less pronounced in the present case. Most importantly, both $T_{cc} \pi $ and 
$T_{cc} \rho $ absorption cross sections are greater than the respective  
production ones. 

There are some differences between the empirical and the QCDSR approach.
For example, some reactions in Figs.~\ref{Fig:CrSec-Abs-QCDSR} and       
~\ref{Fig:CrSec-Prod-QCDSR} have greater magnitudes while others have 
smaller magnitudes when compared to the corresponding ones in             
~\ref{Fig:CrSec-Abs-s} and ~\ref{Fig:CrSec-Prod-s}. This is mainly due to
the fact that in the QCDSR approach each vertex in a given diagram has a
distinct form factor with a specific parametrization according to
Eqs.~(\ref{exp})-(\ref{FII}) and Table  \ref{FFparam}.

\begin{figure}[!ht]                                                            
   \centering                                                                 
   \includegraphics[{width=1.0\linewidth}]{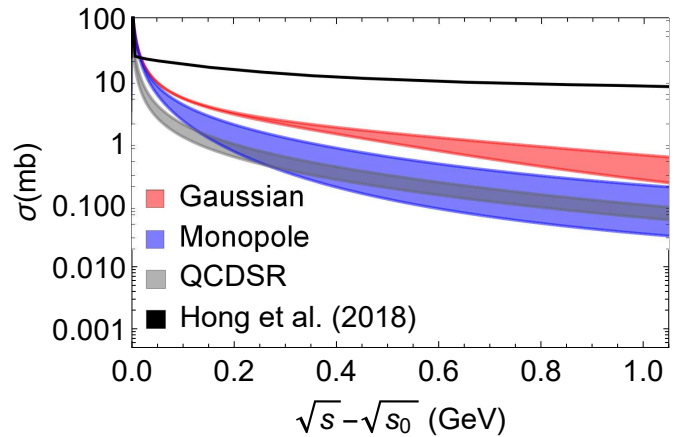}
   \caption{ Sum of the cross sections for the processes obtained in this
     work ($T_{cc} \pi \rightarrow D D + D D^{*} + D^{*} D^{*}$) and in
     Ref.~\cite{Hong:2018mpk} ($T_{cc} \pi \rightarrow D \pi  + D^{*} \pi $ ),
     as a function of the CM energy $\sqrt{s}$ above the threshold energy
     $\sqrt{s_0}$ of each process. The legend denotes the respective functional
     form of form factors used to calculate the cross sections.} 
   \label{comp}
   \end{figure}

Finally, in Fig.~\ref{comp} we compare our results with those obtained in
Ref.~\cite{Hong:2018mpk}. As it  can be seen from the figure, our cross
sections are significantly smaller than those found in \cite{Hong:2018mpk}. 
In comparison to the effective Lagrangian adopted here, in the quasi-free
approximation the $T_{cc}^+$ is ``too easy to destroy''. 
Part of this difference can be attributed to the couplings and uncertainties
associated to the form factors used in each work. 

In Fig.~\ref{comp} we can observe that the results obtained with the empirical 
and the QCDSR form factors are not so different. This is not unexpected, since
the form factors are calibrated with coupling constants, which, in some
vertices, 
were the same and derived from QCDSR. So, in fact, most of the difference in the
cross sections comes from how the form factor decreases with increasing $Q^2$.

In the end, the final results contain  large uncertainties, but they still carry 
valuable information. Despite all uncertainties, it remains true that
$T_{cc}$ absorption is stronger than its creation. This result is not
surprising and a similar dominance of absorption over production was
found in the case of  $J/\psi$, $\Upsilon$, and other multiquark states 
such as the $X(3872)$ and $Z_b$.

Once the vacuum cross sections are known, the next step is to compute the
thermal cross sections, which are convolutions of the vacuum cross sections
with thermal momentum distributions of the colliding particles. In this
approach, the temperature of the hadron gas (which is in the range 100 - 200
MeV) determines the collision energy. When we perform this thermal average,
the kinematical configurations close to the thresholds are highly suppressed.
Hence the strong threshold enhancement (or suppression) observed in all the
figures above have little significance for  applications to heavy ion
collisions.

%

\section{Concluding remarks}

\label{Conclusions}
The purpose of this work has been to study, from an effective framework, the interactions of the doubly charmed tetraquark state $T_{cc} ^+$ with light mesons in the hadron gas phase. Their  absorption and production processes have been computed, deserving special attention to the role of the form factor of the $T_{cc}-D-D^*$  coupling, obtained from QCD sum rules. 

The results suggest sizeable cross sections for 
the considered processes. Moreover,  the $T_{cc}^+$ absorption in a
hadron gas appears to be more important than its production. On the other hand, when compared with the other existing estimate, our approach suggests much smaller $T_{cc} ^+$ absorption cross sections than those of Ref.~\cite{Hong:2018mpk}, based on the quasi-free approximation.

The obtained $T_{cc}^+$ production and absorption cross sections will
be crucial for a comprehensive analysis
of the evolution of the  $T_{cc}^+$ abundance in heavy ion collisions. 
This is another observable useful to shed light on the $T_{cc}^+$ internal 
structure. This study is in progress and we expect to publish it soon.

\begin{acknowledgements}

  The authors would like to thank the Brazilian funding agencies for their
  financial support: CNPq (LMA: contracts 309950/2020-1 and 400546/2016-7), 
  FAPESB (LMA: contract INT0007/2016) and INCT-FNA.

\end{acknowledgements} 




\begin{thebibliography}{99}


\bibitem{LHCb:2021vvq}   R.~Aaij \textit{et al.} [LHCb],
                         [arXiv:2109.01038 [hep-ex]].

\bibitem{LHCb:2021auc}   R.~Aaij \textit{et al.} [LHCb],
                         [arXiv:2109.01056 [hep-ex]].


\bibitem{Gelman:2002wf}  B.~A.~Gelman and S.~Nussinov,
                         Phys. Lett. B \textbf{551}, 296 (2003).


\bibitem{Janc:2004qn}    D.~Janc and M.~Rosina,
                         Few Body Syst. \textbf{35}, 175 (2004). 

\bibitem{Vijande:2003ki} J.~Vijande, F.~Fernandez, A.~Valcarce and
                         B.~Silvestre-Brac,
                         Eur. Phys. J. A \textbf{19}, 383 (2004). 


\bibitem{Navarra:2007yw} F.~S.~Navarra, M.~Nielsen and S.~H.~Lee,
                         Phys. Lett. B \textbf{649}, 166 (2007). 

\bibitem{Vijande:2007rf} J.~Vijande, E.~Weissman, A.~Valcarce and N.~Barnea,
                         Phys. Rev. D \textbf{76}, 094027 (2007). 

\bibitem{Ebert:2007rn}   D.~Ebert, R.~N.~Faustov, V.~O.~Galkin and W.~Lucha,
                         Phys. Rev. D \textbf{76}, 114015 (2007). 

\bibitem{Lee:2009rt}     S.~H.~Lee and S.~Yasui,
                         Eur. Phys. J. C \textbf{64}, 283 (2009).


\bibitem{Yang:2009zzp}   Y.~Yang, C.~Deng, J.~Ping and T.~Goldman,
                         Phys. Rev. D \textbf{80}, 114023 (2009). 

\bibitem{Hong:2018mpk}   J.~Hong, S.~Cho, T.~Song and S.~H.~Lee,
                         Phys. Rev. C \textbf{98},  014913 (2018).

                         
\bibitem{Hudspith:2020tdf} R.~J.~Hudspith, B.~Colquhoun, A.~Francis,
                           R.~Lewis and K.~Maltman,
                           Phys. Rev. D \textbf{102}, 114506 (2020). 

\bibitem{Cheng:2020wxa}    J.~B.~Cheng, S.~Y.~Li, Y.~R.~Liu, Z.~G.~Si
                           and T.~Yao,
                           Chin. Phys. C \textbf{45}, 043102 (2021). 

\bibitem{Qin:2020zlg}      Q.~Qin, Y.~F.~Shen and F.~S.~Yu,
                           Chin. Phys. C \textbf{45}, 103106 (2021). 

\bibitem{Agaev:2021vur}    S.~S.~Agaev, K.~Azizi and H.~Sundu,
                           [arXiv:2108.00188 [hep-ph]].


\bibitem{Dong:2021bvy}     X.~K.~Dong, F.~K.~Guo and B.~S.~Zou,
                           [arXiv:2108.02673 [hep-ph]].


\bibitem{Huang:2021urd}    Y.~Huang, H.~Q.~Zhu, L.~S.~Geng and R.~Wang,
                           Phys. Rev. D \textbf{104},  116008 (2021).

\bibitem{Li:2021zbw}       N.~Li, Z.~F.~Sun, X.~Liu and S.~L.~Zhu,
                           Chin. Phys. Lett. \textbf{38}, 092001 (2021). 

                           
\bibitem{Ren:2021dsi}      H.~Ren, F.~Wu and R.~Zhu,
                           [arXiv:2109.02531 [hep-ph]].


\bibitem{Xin:2021wcr}      Q.~Xin and Z.~G.~Wang,
                           [arXiv:2108.12597 [hep-ph]].




\bibitem{Yang:2021zhe}
G.~Yang, J.~Ping and J.~Segovia,
[arXiv:2109.04311 [hep-ph]].


\bibitem{Meng:2021jnw} L.~Meng, G.~J.~Wang, B.~Wang and S.~L.~Zhu,
                       Phys. Rev. D \textbf{104}, 051502 (2021). 

\bibitem{Ling:2021bir}  X.~Z.~Ling, M.~Z.~Liu, L.~S.~Geng,
                        E.~Wang and J.~J.~Xie,
                        [arXiv:2108.00947 [hep-ph]].


\bibitem{Fleming:2021wmk}  S.~Fleming, R.~Hodges and T.~Mehen, 
                           [arXiv:2109.02188 [hep-ph]].


\bibitem{Jin:2021cxj} Y.~Jin, S.~Y.~Li, Y.~R.~Liu, Q.~Qin, Z.~G.~Si and
                      F.~S.~Yu, [arXiv:2109.05678 [hep-ph]].


\bibitem{Azizi:2021aib}  K.~Azizi and U.~\"Ozdem, [arXiv:2109.02390 [hep-ph]].

\bibitem{Hu:2021gdg}
Y.~Hu, J.~Liao, E.~Wang, Q.~Wang, H.~Xing and H.~Zhang,
[arXiv:2109.07733 [hep-ph]].


\bibitem{Chen:2007zp}  L.~W.~Chen, C.~M.~Ko, W.~Liu and M.~Nielsen,
                       Phys. Rev. C \textbf{76}, 014906  (2007);
                       F.~S.~Navarra, M.~Nielsen, M.~E.~Bracco,
                       M.~Chiapparini and C.~L.~Schat,
                       Phys. Lett. B \textbf{489}, 319 (2000);
                       F.~S.~Navarra, M.~Nielsen and M.~E.~Bracco,
                       Phys. Rev. D \textbf{65}, 037502 (2002). 


\bibitem{ChoLee1}      S.~Cho and S.~H.~Lee,
                       Phys. Rev. C {\bf 88}, 054901 (2013); 
                       M.~E.~Bracco, M.~Chiapparini, F.~S.~Navarra
                       and M.~Nielsen, Phys. Lett. B \textbf{659}, 559 (2008).
                       

\bibitem{XProd1}       A. Martinez Torres, K. P. Khemchandani, F. S. Navarra, 
                       M. Nielsen and L. M. Abreu, 
                       Phys. Rev. D {\bf 90}, 114023  (2014);
                       A. Martinez Torres, K. P. Khemchandani, F. S. Navarra, 
                       M. Nielsen and L. M. Abreu, 
                       Acta Phys. Pol. B Proc. Supp. {\bf 8}, 247 (2015). 


\bibitem{XProd2}     L. M. Abreu, K. P. Khemchandani, A. Martinez Torres, 
                     F. S. Navarra and M. Nielsen, 
                     Phys. Lett. B {\bf 761}, 303 (2016).

\bibitem{UFBaUSP1} L. M. Abreu, K. P. Khemchandani, A. Martinez Torres, 
                   F. S. Navarra, M. Nielsen and A. L. Vasconcellos, 
                   Phys. Rev. D {\bf 95}, 096002 (2017). 


\bibitem{MartinezTorres:2017eio}   A.~Mart\'\i{}nez Torres,
                   K.~P.~Khemchandani, L.~M.~Abreu, F.~S.~Navarra
                   and M.~Nielsen, 
                   Phys. Rev. D \textbf{97}, 056001 (2018). 


\bibitem{Abreu:2017cof} L.~M.~Abreu, K.~P.~Khemchandani,
                        A.~Mart\'\i{}nez Torres, F.~S.~Navarra and M.~Nielsen,
                        Phys. Rev. C \textbf{97}, 044902 (2018). 

                        
\bibitem{Abreu:2018mnc} L.~M.~Abreu, F.~S.~Navarra and M.~Nielsen,
                        Phys. Rev. C \textbf{101}, 014906 (2020).

\bibitem{LeRoux:2021adw} C.~Le Roux, F.~S.~Navarra and L.~M.~Abreu,
                         Phys. Lett. B \textbf{817}, 136284 (2021). 


\bibitem{shm1}           A.~Andronic, P.~Braun-Munzinger,
                         M.~K.~K\"ohler, K.~Redlich and J.~Stachel,
                         Phys. Lett. B \textbf{797}, 134836 (2019).

\bibitem{shm2}           A.~Andronic, P.~Braun-Munzinger, M.~K.~K\"ohler,
                         A.~Mazeliauskas, K.~Redlich, J.~Stachel
                         and V.~Vislavicius,
                         JHEP \textbf{07}, 035 (2021).

                         
\bibitem{psipi-oh}   S. G. Matinyan and B. M\"{u}ller,
                     Phys. Rev. C {\bf 58}, 2994 (1998);
                     Y. Oh, T. Song and S. H. Lee,
                     Phys. Rev. C {\bf 63}, 034901 (2001);
                     F.~Carvalho, F.~O.~Duraes, F.~S.~Navarra and M.~Nielsen,
                     Phys. Rev. C \textbf{72}, 024902 (2005);  
                     B.~Osorio Rodrigues, M.~E.~Bracco, M.~Nielsen
                     and F.~S.~Navarra, Nucl. Phys. A \textbf{852}, 127 (2011).

\bibitem{svz}    M.A. Shifman, A.I. and Vainshtein and V.I. Zakharov,
                 Nucl. Phys. B {\bf 147}, 385 (1979).

\bibitem{rry}    L.J. Reinders, H. Rubinstein and S. Yazaki,
                 Phys. Rept. {\bf 127}, 1 (1985).


\bibitem{bcnn11} M.E.~Bracco, M.~Chiapparini, F.S.~Navarra and M.~Nielsen,
                 Prog.\ Part.\ Nucl.\ Phys.\  {\bf 67}, 1019 (2012).

\bibitem{kho95}  V.~M.~Belyaev, V.~M.~Braun, A.~Khodjamirian and
                 R.~Ruckl, Phys. Rev. D \textbf{51}, 6177 (1995).
                 
\bibitem{kho21}  A.~Khodjamirian, B.~Meli\'c, Y.~M.~Wang and Y.~B.~Wei,
                 JHEP \textbf{03}, 016 (2021).                                  
  
\bibitem{Dias:2013xfa} J.~M.~Dias, F.~S.~Navarra, M.~Nielsen and C.~M.~Zanetti,
                       Phys. Rev. D \textbf{88}, 016004 (2013). 

                       
\bibitem{io1} B.L. Ioffe and A.V. Smilga, { Nucl. Phys.} B {\bf 232},  109
(1984).

\bibitem{Zyla:2020zbs}  P.~A.~Zyla \textit{et al.} [Particle Data Group],
                        PTEP \textbf{2020},  083C01 (2020).

  
                            


\end{thebibliography}
\end{document}